\documentclass{aa}  

\usepackage{graphicx}
\usepackage{txfonts}
\usepackage{natbib}
\usepackage{lscape}

\begin{document}
 
\title{
Gravitational lensing and dynamics in \object{SL2S\,J02140-0535}:\\
Probing the mass out to large radius}
   \titlerunning{Strong Lensing by Galaxy Groups}
   \authorrunning{Verdugo et~al.}
	\subtitle{}
   \author{T. Verdugo,\inst{1}
                V. Motta,\inst{1}
                R. P. Mu\~{n}oz,\inst{1}
                M. Limousin,\inst{2,3}
                R. Cabanac,\inst{4}
                J. Richard\inst{5}
                      \thanks{Based on observations obtained with MegaPrime/MegaCam, a joint project of CFHT and CEA/DAPNIA, at the Canada-France-Hawaii Telescope (CFHT) which is operated by the National Research Council (NRC) of Canada, the Institut National des Sciences de l'Univers of the center National de la Recherche Scientifique (CNRS) of France, and the University of Hawaii. This work is based in part on data products produced at TERAPIX
and the Canadian Astronomy Data center as part of the Canada-France-Hawaii Telescope Legacy Survey, a collaborative project of NRC and CNRS. Also based on HST data as well as Keck (LRIS) and VLT (FORS\,2) data.}
       }
   \offprints{tverdugo$@$dfa.uv.cl}

   \institute{Universidad de Valpara\'{\i}so, Departamento de F\'{\i}sica y Astronom\'{\i}a, Avenida Gran Breta\~{n}a 1111, Valpara\'{\i}so, Chile
                 \and
                 Laboratoire d'Astrophysique de Marseille, Universit\'e de Provence,\\ CNRS, 38 rue Fr\'ed\'eric Joliot-Curie, F-13388 Marseille Cedex
13, France
                \and
        Dark Cosmology center, Niels Bohr Institute, University of Copenhagen,
       Juliane Marie Vej 30, 2100 Copenhagen, Denmark %\\
           \and
	Laboratoire d'Astrophysique de Toulouse-Tarbes, Universit\'e de Toulouse, CNRS,
     57 Avenue d'Azereix, 65 000 Tarbes, France
           \and
           Durham University, Physics and Astronomy Department, South Road, Durham DH3 1LE, UK
           }

 %\date{Accepted for publication }
 \date{Preprint online version:}

 \abstract
  {Studying the density profiles of galaxy groups offers an important insight into the formation and evolution of the structures in the universe, since galaxy groups bridge the gap between single galaxies and massive clusters.}
  {We aim to probe the mass of \object{SL2S\,J02140-0535}, a galaxy group at $z=0.44$
from the Strong Lensing Legacy Survey (SL2S), which has uncovered this new
population of group-scale strong lenses.
 }
  {We combine strong lensing modeling and dynamical constraints.
The strong lensing analysis is based on multi-band HST/ACS observations
exhibiting strong lensing features that we have followed-up
spectroscopically with VLT/FORS2. To constrain the scale radius of an NFW mass profile that cannot be constrained by strong lensing, we propose a new method by 
taking advantage of the large-scale dynamical information provided by
VLT/FORS2 and KECK/LRIS spectroscopy of group members.  
}
  {In constrast to other authors, we show that the observed lensing features
in \object{SL2S\,J02140-0535}  belong to different background
sources: one at  $z$= 1.7 $\pm$ 0.1 (photometric redshift) produces
three images, while the other at  $z$ = 1.023 $\pm$ 0.001 (spectroscopic redshift) has only a single image. Our unimodal NFW mass model reproduces these images very well.
It is characterized by a concentration parameter $c_{200}$ = 6.0 $\pm$ 0.6,
which is slightly greater than the value expected from $\Lambda$CDM simulations for a mass of M$_{200}$ $\approx$ 1 $\times$ 10$^{14}$ M$_{\sun}$.
The spectroscopic analysis of group members also reveals a unimodal structure that exhibits no evidence of merging.
The position angle of the halo is $\theta$ = 111.6 $\pm$ 0.2, which 
agrees with the direction defined by the luminosity contours.
We compare our dynamic mass estimate with an independent weak-lensing based mass estimate finding that both are consistent.  %, arguing for a relaxed galaxy group.  
}
{Our combined lensing and dynamical analysis of \object{SL2S\,J02140-0535} demonstrates the importance of spectroscopic information in reliably identifying the lensing features. Our findings argue that the system is a relaxed, massive galaxy group where mass is traced by light. This work shows a potentially useful method for constraining large-scale properties inaccessible to strong lensing, such as the scale radius of the NFW profile.
}

  \keywords{Gravitational lensing: strong lensing --
               Galaxies: groups  --
              Galaxies: groups: individual (\object{SL2S\,J02140-0535})
}

  % \titlerunning{NFW vs. PIEMD}
   \titlerunning{Lensing and dynamics in \object{SL2S\,J02140-0535}}
  \authorrunning{Tom\'as Verdugo et al.}
  \maketitle
%
%________________________________________________________________

 %  \maketitle
%________________________________________________________________

 %%%%%%%%%%%%%%%%%%%%%%%%%%%%%%%%%%%
\begin{figure*} \begin{center}
\includegraphics[scale=0.567]{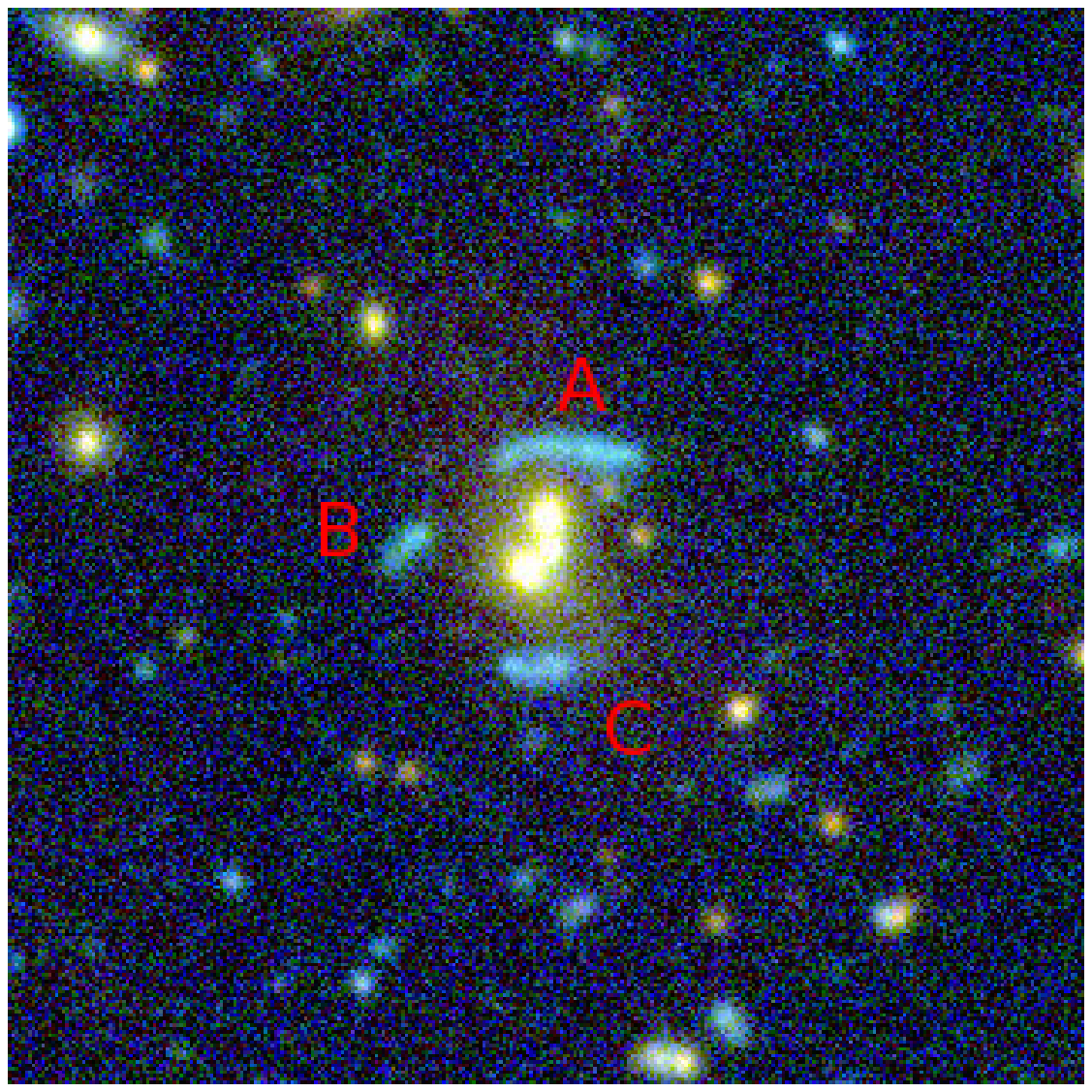}
\includegraphics[scale=1.1435]{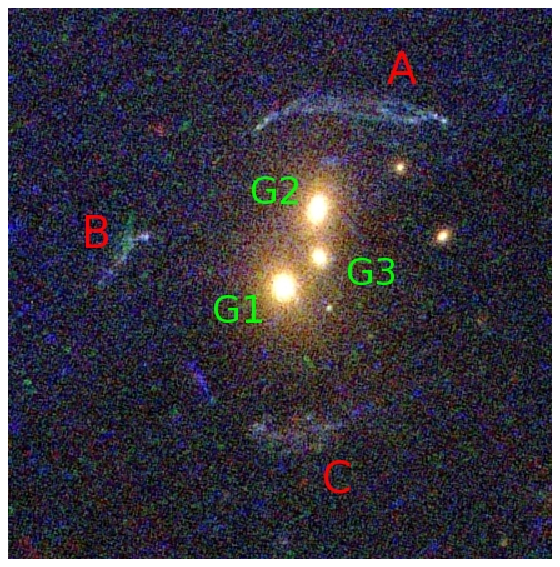}\\
\includegraphics[scale=0.8]{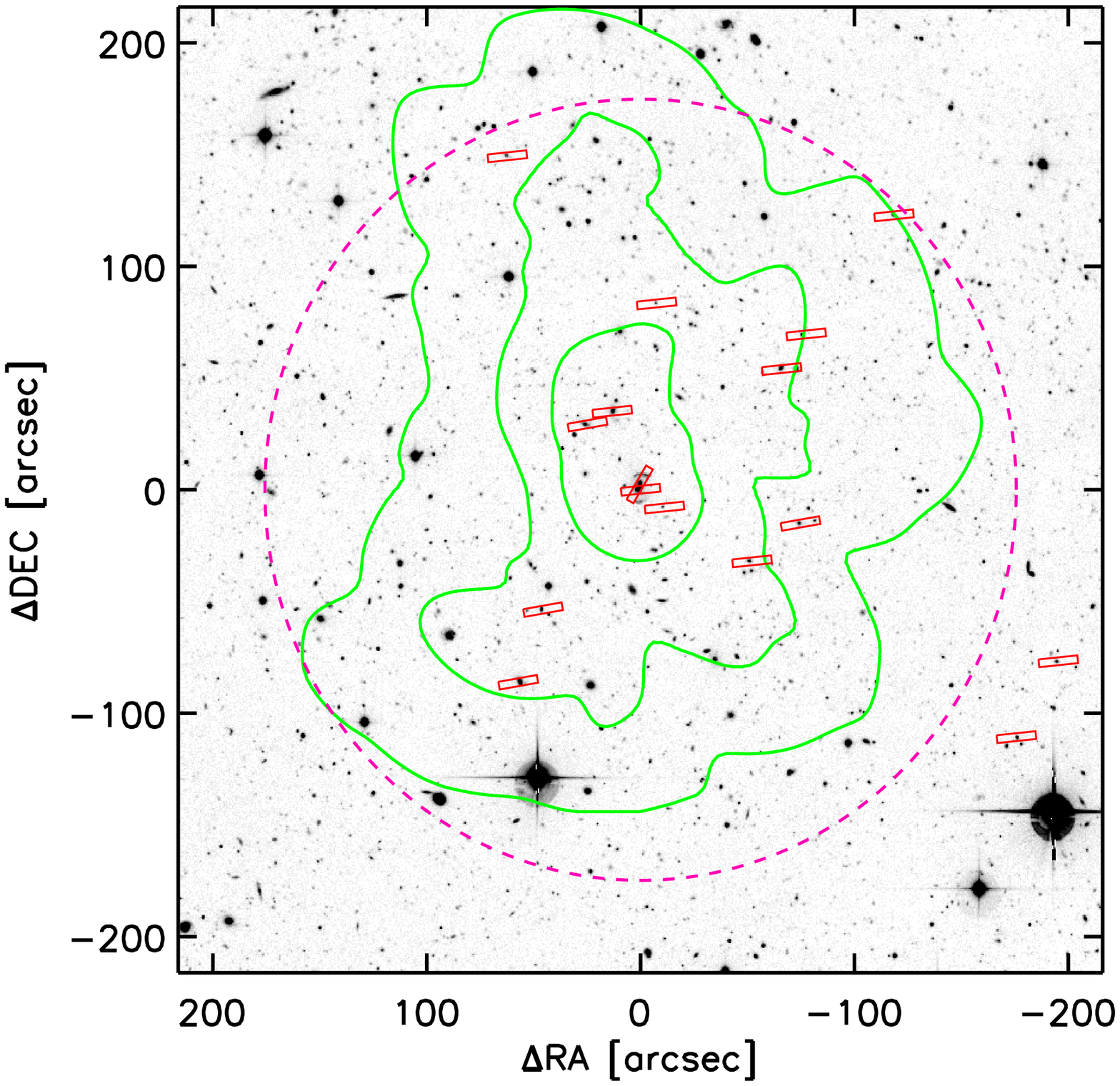}
\caption{The group SL2S\,J02140-0535 at $z_{\rm spec}=0.44$.
\emph{Upper left:} composite CFHTLS $g,r,i$ color image (1 square arcminute = 341 $\times$ 341 kpc$^{2}$).
\emph{Upper right:} composite HST/ACS F814, F606, F475 color image ($22\arcsec \times 22\arcsec$ = 125 $\times$ 125 kpc$^{2}$).
Note how the space-based image significantly improves the
identification of multiple images and heightens the resolution allowing us to detect the substructure. \emph{Lower:} CFHTLS $i$ band ($7.2'\,\times\,7.2'$ = 2.46 $\times$ 2.46 Mpc$^{2}$) centered
on the strong lensing system.
We draw in green the luminosity contours (corrected for passive evolution) corresponding
to 10$^{6}$, 3 $\times$ 10$^{6}$, and 10$^{7}$ L$_{\sun}$\,kpc$^{-2}$, respectively. Red rectangles show the location of the slits used for spectroscopy for the 16 confirmed members of the group and the magenta circle denotes a 1 Mpc distance from the center of the group.
}
\label{presentlens} \end{center} \end{figure*}
%%%%%%%%%%%%%%%%%%%%%%%%%%%%%%%%%%%%%%%%

%%%%%%%%%%%%%%%%%%%%%%%%%%%%%%%%%%%%%%%%
\begin{table*}
\caption{Photometric and spectroscopic data of the galaxies and the arcs.}
\label{tbl-1} % is used to refer this table in the text
\centering % used for centering table
\begin{tabular}{lccccccccc}
\hline\hline % inserts double horizontal lines
\\
 ID Name  &  X   &   Y   &
\multicolumn{1}{c}{$u^{*}-g'$}  &
\multicolumn{1}{c}{$g'-r'$}  &
\multicolumn{1}{c}{$r'-i'$}  &
\multicolumn{1}{c}{$i'-z'$}  &  $\epsilon$
&    \multicolumn{1}{c}{$\theta_0$} &  z    \\

  &  [\arcsec]  &    [\arcsec]  & & &  &
 &  
&  $[^{\circ}]$   &      \\

\\
\hline % inserts single horizontal line
\\
$G1$ & $0.00$ & $0.00$ & $2.35\pm0.11  $ & $1.61\pm0.02$ & $0.86\pm0.01$ & $0.36\pm0.01$ & $0.158$ & $-41.0$ & $0.4446\pm0.0002$ \\
$G2$ & $1.16$ & $3.09$ & $2.99\pm0.23$ &$1.64\pm0.03$ & $0.87\pm0.01$ & $0.37\pm0.01$  &  $0.424$ & $\phantom{-}64.2$ & $0.4449\pm0.0003$ \\
$G3$ & $1.37$ & $1.25$ & $1.81\pm0.15$  & $1.47\pm0.06$  & $0.88\pm0.04$ & $0.33\pm0.03$ & $0.177$ & $\phantom{-0}18.26$ & --- \\
$A$ & --- & --- & $0.73\pm0.09$  & $0.30\pm0.09$ & $0.47\pm0.10$ & $0.31\pm0.11$ &  --- & --- & [$1.7\pm0.1$] \\
$B$ & --- & --- & $0.60\pm0.11  $&  $0.07\pm0.11$ & $0.39\pm0.14$  & $0.36\pm0.14$  &  --- & --- &[$1.7\pm0.1$]\\
$C$ & --- & --- & $0.18\pm0.09$  & $0.58\pm0.09$  & $0.60\pm0.09$  & $0.41\pm0.09$ &  --- & --- & $1.023\pm 0.001$ \\
\\
\hline %inserts single line
\end{tabular}
\tablefoot{
Column (1) lists the identification for each object as in Fig.~\ref{presentlens}. Columns (2) and (3) list the relative positions measured with respect to the brightest group galaxy (BGG) with coordinates $\alpha$ = 02:14:08:097 and $\delta$ = -05:35:33.77. Columns (4-7) list the CFHTLS colors. Columns (8) and (9) show the geometric parameters: ellipticity $\epsilon$ and position angle $\theta_0$, which gives the orientation of the semi-major axis from the horizontal line in the image, measured counterclockwise. In the last column we list the redshift; values in square brackets are photometric redshifts.
}
\end{table*}
%%%%%%%%%%%%%%%%%%%%%%%%%%%%%%%%%%

\section{Introduction}

In the context of structure formation and evolution, galaxy groups play a key role,
spanning the regime between individual galaxies and galaxy clusters.
Therefore, detailed studies of this intermediate regime of the mass spectrum are relevant.
Galaxy groups have been mostly studied using X-ray and optical tracers, as well as numerical
simulations \citep[\emph{e.g.}][and references therein]{group1,group2,group3,group4,group5,group6,mamongroup,group7,group8,group9,fabio,elena05,jespergroup}.
Only recently, thanks to the unsurpassed combined depth, area, and image quality of the Canada France Hawaii Telescope Legacy Survey (CFHTLS)\footnote{\tt http://www.cfht.hawaii.edu/Science/CFHLS/}, a new population of lenses has been uncovered, with Einstein radii between $\sim 3\arcsec$ and $\sim 8\arcsec$, generated by galaxy-group-scale dark-matter haloes \citep[this subsample of the CFHTLS forms the Strong Lensing Legacy Survey\footnote{\tt http://www-sl2s.iap.fr/} (SL2S),][]{sl2s}.
This new population effectively bridges the gap between single galaxies and massive clusters, opening a new window of exploration in the mass spectrum. The first representative sample of these lenses was presented by \citet{paperI}.
Since then, a detailed investigation of some groups has been carried out: SL2S\,J02176-0513 \citep{hong09} and SL2S\,J08544-0121 \citep{my0854}.  In this paper, we present a detailed analysis of \object{SL2S\,J02140-0535} at $z=0.44$, combining the
dynamics of the group members, and both the strong and weak gravitational lensing.  

Some researchers have demonstrated the advantages of combining the different lensing regimes with the constraints that come from the stellar kinematics of the central brightest cluster galaxy (BCG). For example, \citet{sand02,sand04} combined strong gravitational lensing with the velocity dispersion profile of the BCG and found that the models favored a shallower slopes in the NFW profiles.  \citet{new09} following the above works and adding weak-lensing analysis, sampled the dark matter profile over three decades in radius in the lensing cluster Abell\,611. Despite all these works, on group scales the efforts in that direction are just beginning. First, because strong lensing in groups is a relatively new field of research, and second, owing to observational limitations, i.e., there are fewer lensing arcs in groups and detecting weak lensing by groups of galaxies is challenging.  \citet{mck10} probed the inner part of a group-scale halo using strong gravitational lensing and stellar kinematics, and  \citet{tha10} showed that is possible to combine strong lensing and galaxy dynamics to characterize the mass distribution and the mass-to-light ratios of galaxy groups.

 \object{SL2S\,J02140-0535} was previously studied by \citet{alardalone} using a perturbative reconstruction of the gravitational lens. He found that the local shape of the potential and density of the lens, inferred from the perturbative solution, reveals the existence of an independent dark component that does not follow light.
He argued that this particular shape of the dark halo is due to the merging of cold dark matter halos.   
Given these findings, the group properties represent a challenge, and to describe its mass distribution and perform a detailed study of the object is of significant worth and deserves attention. Our first aim in this work is to build a strong lensing mass model for \object{SL2S\,J02140-0535} using lensed features observed in HST/ACS data combined with ground-based spectroscopy. The second is to combine the strong lensing constraints with the dynamics of the groups members to study the mass profile out to the scale radius. Finally, we wish to compare the masses obtained with the three methods, namely, strong lensing, weak lensing, and group dynamics, to shed some light on \object{SL2S\,J02140-0535}.

The paper is organized as follows. In Sect.\,\ref{data}, we describe the observational data
images and spectroscopy and present the multiple image identification. In Sect.\,\ref{dynamics}, we define how the dynamical constraints (on galaxy and group scales) enter into our strong lensing model. We also depict the profiles for both, the group and the galaxy-scale mass components. In Sect.\,\ref{model}, we present the strong lensing as well as the weak lensing analysis of \object{SL2S\,J02140-0535} and the main results of the work. In Sect.\,\ref{discuss}, we summarize and discuss our results. Finally in Sect.\,\ref{conclus}, we present the conclusions. All our results are scaled to a flat, $\Lambda$CDM cosmology with $\Omega_{\rm{M}} = 0.3, \Omega_\Lambda = 0.7$ and a Hubble constant \textsc{H}$_0 = 70$
km\,s$^{-1}$ Mpc$^{-1}$. All images are aligned with WCS coordinates, i.e. north is up, east is left. Magnitudes are given in the AB system.

\section{Observations and multiple images identification}\label{data}

\object{SL2S\,J02140-0535} has been imaged by both ground and space based telescopes. From the ground, this group was observed in five filters ($u^*,g',r',i',z'$) as part of the CFHTLS using the wide field imager \textsc{MegaPrime}, which covers $\sim$\,1 square degree on the sky, and a pixel size of 0.186$\arcsec$. From space, the lens was followed-up with the \emph{Hubble Space Telescope} (HST). Observations were performed in snapshot mode (C\,15, P.I. Kneib) in three bands with the ACS camera (F814, F606, and F475). Moreover,
a spectroscopic follow-up of the arcs and the group members have been carried out from VLT and Keck.

\subsection{Imaging}

The strong lensing deflector in this group is populated by three galaxies (see Fig.~\ref{presentlens} top left). The lensed images consist of three arcs surrounding the deflectors: one arc (labeled A) situated north of the deflector composed by two merging images, a second arc in the east (labeled B), and a third one (labeled C) in the south. In this ground-based image, we see a possible system consisting of ABC images that seem to have compatible colors. In Fig.~\ref{presentlens} (top right), we show a color composition from the HST-ACS camera for the central part of the group. Since in the HST image, arc C exhibits different color from both A and B, is it possible that this arc is not associated with the proposed multiple-image system? This scenario is supported by the substructure in arcs A and B that are not present in arc C (see top right of Fig.~\ref{presentlens}). Given that \citet{alardalone} published an analysis based on this southern image being part of the multiply imaged system, special care will be used to properly assess the multiple image identification, which is crucial for the analysis.

The photometry for the central galaxies labeled G1, G2, and G3 (see top right of Fig.~\ref{presentlens}),  and for the arcs was performed in all CFHT bands with the IRAF package apphot. For the galaxies, the magnitudes were derived by measuring fluxes inside a fixed circular aperture of 12 pixels ($\sim$ 2.22$\arcsec$) for galaxies G1 and G2, and 6 pixels ($\sim$ 1.12 $\arcsec$) in galaxy G3.  For the arcs, we employed polygonal apertures to obtain more accurate measurements. The vertices of the polygons for each arc were determined using the IRAF task polymark, and the magnitudes inside these apertures were calculated using the IRAF task polyphot.

The general properties of the central galaxies as well as the arcs are presented in Table~\ref{tbl-1} . Column (1) lists the identification for each object. Columns (2) and (3) list the relative positions measured with respect to the brightest group galaxy (BGG) with coordinates $\alpha$ = 02:14:08:097 and $\delta$ = -05:35:33.77. Columns (4-7) list the colors $u^*-g'$, $g'-r'$, $r'-i'$ and $i'-z'$. Columns (8) and (9) show the geometric parameters derived from SExtractor \citep{sextractor}, i.e., ellipticity $\epsilon$ and position angle $\theta_0$, which imply the orientation of 
the semi-major axis from the horizontal line in the image, measured counterclockwise. Finally, column (10) lists the redshifts (see the next subsections). We note in Table~\ref{tbl-1} that the colors $u^*-g'$ and $g'-r'$ for arc C 
differ significantly from the colors in A and B, which have almost similar values (within the errors).

%%%%%%%%%%%%%%%%%%%%%%%%%%
\begin{figure}[h!]
\begin{center}
\includegraphics[scale=0.45]{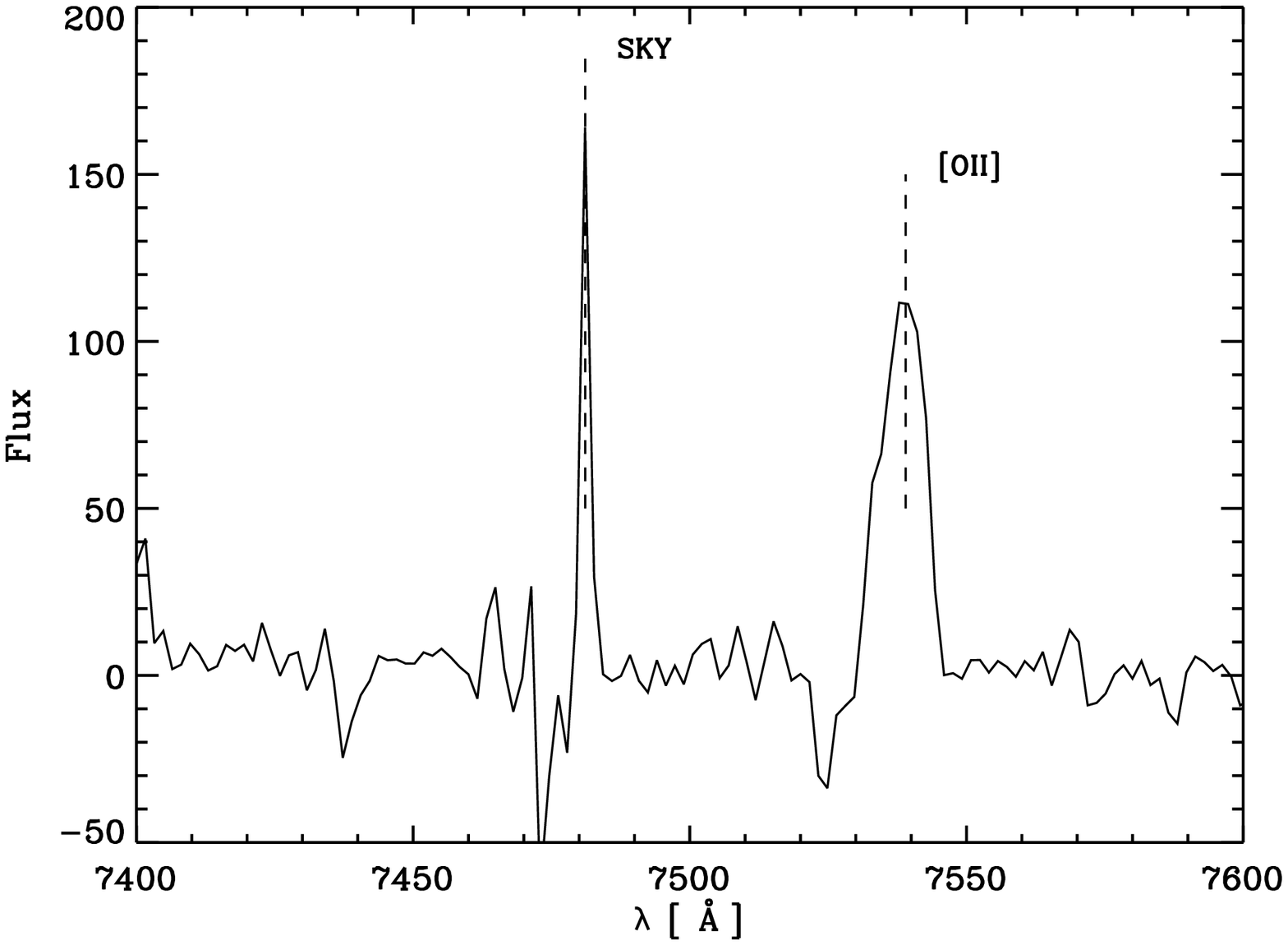}
\includegraphics[scale=0.22]{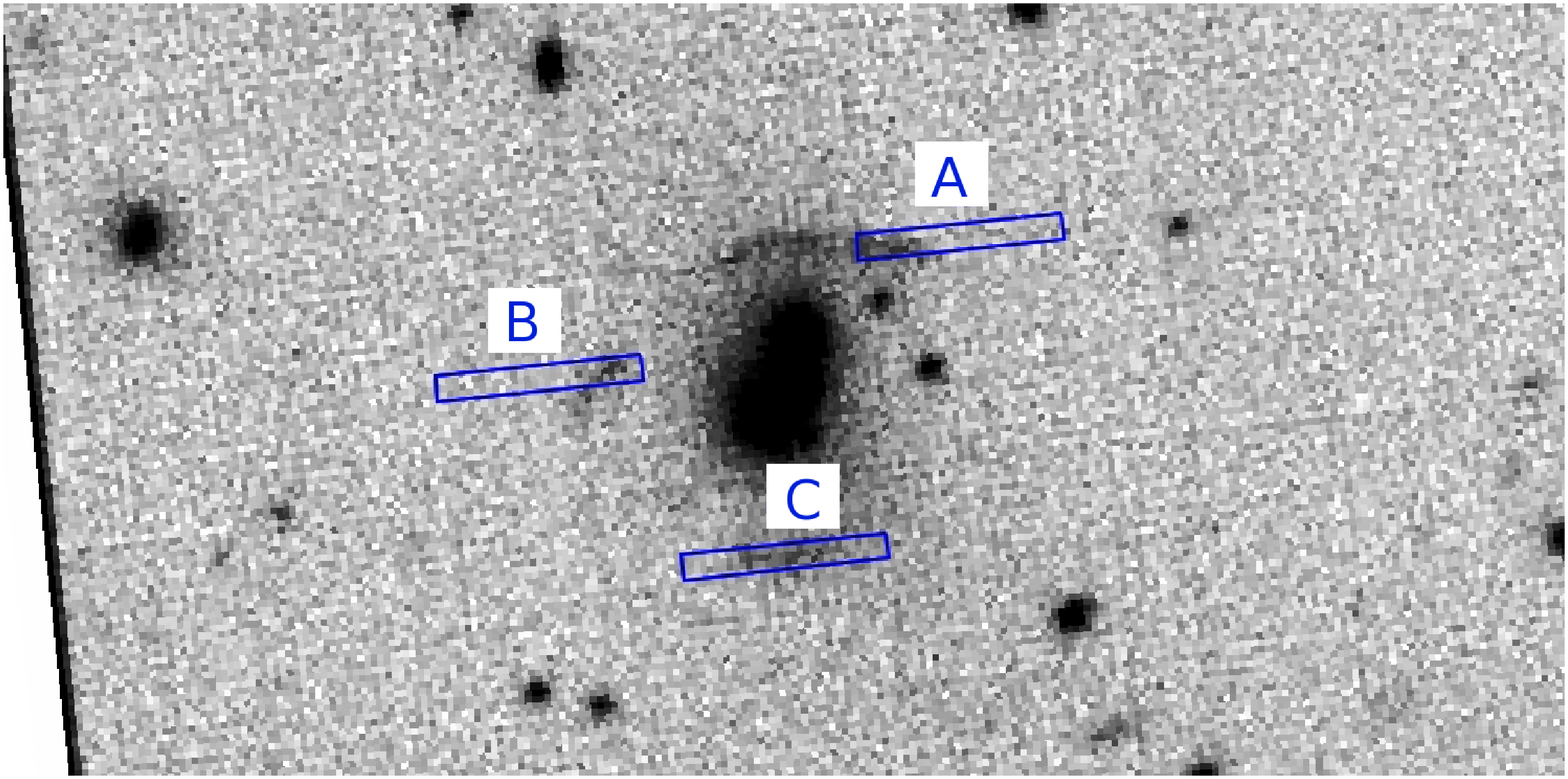}
\includegraphics[scale=0.2615]{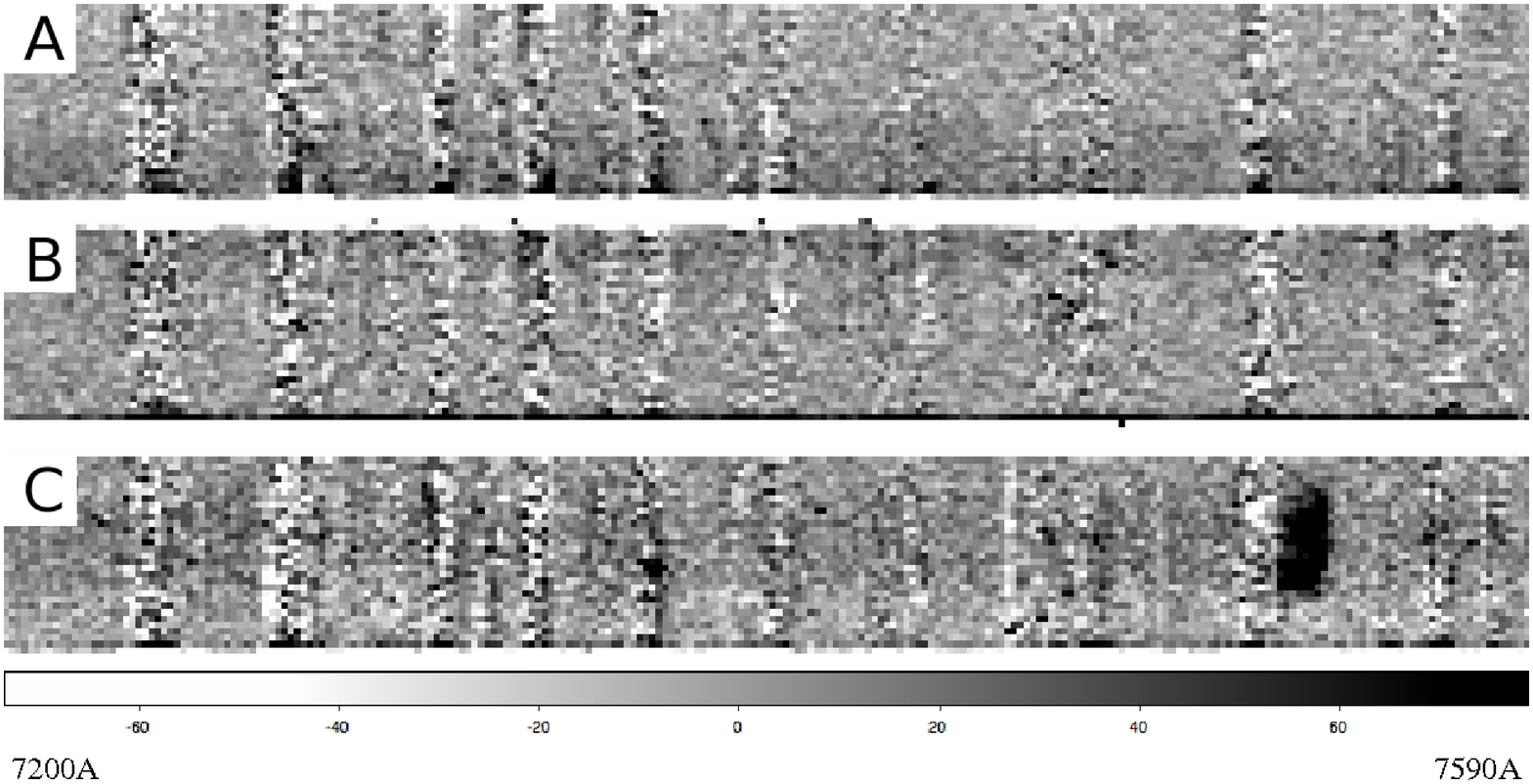}
\caption{\textit{Top panel.} Spectrum of arc C. It shows a strong feature probably associated with [OII]$\lambda$3727. \textit{Middle panel.} VLT pre-image of the central part of \object{SL2S\,J02140-0535} with the slits positions. \textit{Bottom panel.}  The 2D spectra of arcs A, B, and C in the same wavelength region. Note the extended emission line in C spectrum. %The green rectangle delineates the spatial extension of the feature in the 2D spectrum; as we can appreciate in the pre-image, the rectangle covers the whole arc.
}
\label{spectro}
\end{center}
\end{figure}
%%%%%%%%%%%%%%%%%%%%%%%%%%

\subsection{Spectroscopy}
Two spectroscopic runs have been performed in the field of SL2S\,J02140-0535.
Using LRIS \citep{lris} on Keck (dichroic 680 and 560 nm, exposure 600s and 1200s with a slit of 1.5$\arcsec$), we measured the redshifts of one of the brightest galaxies populating the strong lensing deflector (G2) and we found $z$ = 0.4449 $\pm$ 0.0003. 

We also used FORS\,2 on VLT (P.I. V. Motta) with a medium resolution grism (GRIS\,600RI) to target both group members 
and the strongly lensed features. Two masks were used and two exposures of
1400s were made with each mask.
We selected the targets (other than strongly lensed features) as a two step process.
First we choose galaxies whose $g-i$ color is within 0.15 of the $g-i$ color of the
two brightest galaxies populating the strong lensing deflector \citep[see,][]{paperI}.
Since the field was too crowded in the center but not fully sampled at larger distances, we
also selected all bright galaxies in the outer field regardless of their color.
Fig.~\ref{presentlens} shows the slit configuration for all confirmed group members.
Full details of the spectroscopy of the group members will be presented in a forthcoming
publication (Mu\~noz et~al., in prep.). Here we concentrate only on the spectroscopy of
the lensed features.

No spectroscopic features were found in arcs A and B.
On the other hand, the spectra of feature C reveals a strong emission line 
at 7538.4\,\AA. As we can appreciate in Fig.~\ref{spectro}, the emission line has a spatial extension that covers the whole arc (although the slits do not completely cover arcs A and B, they cover the brightest parts, and in arc B only a small fraction is outside). From this analysis, two possible scenarios emerge: system AB and arc C come from two different sources, or belong to the same source but the fluxes in system AB are not  strong enough to detect the line. From the magnitudes in $u^{*}$ band, it is possible to measure the flux ratios for the arcs, $f_B$/$f_C$ $\sim$ 0.5 and $f_A$/$f_C$ $\sim$ 2.5 (we used half of the flux in arc A to make the calculations). Assuming that the magnification is not very different in the arcs and they come from one single source, if the emission line is detected in arc C then it should be detected in arc A. It should even be detected in arc B, since the flux is only half of the flux in C and the feature is strong enough (see top and bottom panel of Fig.~\ref{spectro}). That this is not detected, excludes the second scenario and supports the conclusion that system AB and arc C are produced by different sources.

%%%%%%%%%%%%%%%%%%%%
\begin{figure}[h!]
\begin{center}
\includegraphics[scale=0.4]{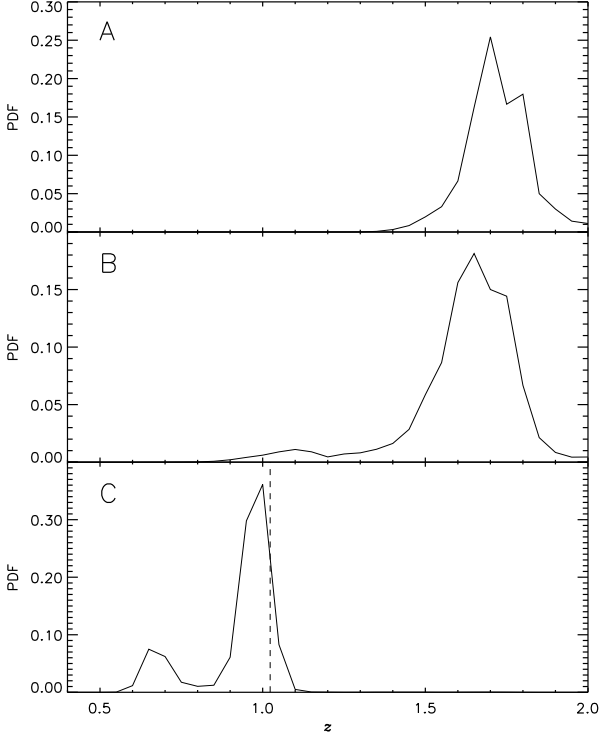}
\caption{Output photometric redshift PDF (see text) for arcs labeled A, B, and C (top to botton). The dashed vertical line corresponds to the spectroscopic redshift inferred assuming the emission line
comes from [OII]$\lambda$3727.}
\label{zphot}
\end{center}\end{figure}
%%%%%%%%%%%%%%%%%%%%

The line detected in C may correspond to either [OII]$\lambda$3727 at $z \sim 1$ or Ly$\alpha$ at $z \sim 5$. However, the low redshift solution is much more likely, and as we discuss below, the photometric redshift estimate supports this conclusion. Thus, assuming emission from [OII]$\lambda$3727 and applying a Gaussian fitting, we obtain  $z_{spec}$ = 1.023 $\pm$ 0.001. This line coming from the whole arc C is inconsistent with the model of \object{SL2S\,J02140-0535} presented in  \citet{alardalone}. He assumed that arcs A, B, and C belong to the same source and he argued that the differences in colors observed in the HST-ACS images were related to the crossing of the caustic line by part of the source. His model predicts that one half of image C comes from the same part of the source as images A and B, but the other half corresponds to an area of the source that has no counterpart in the large arcs (A and B). The lack of spectroscopic features in arcs A and B does not support this scenario.

As a matter of fact, it is known that arcs produced by one single source could have different colors if the source is located over a caustic line. For example, in the case of a
spiral galaxy for which the disk (but not the bulge) is located over a caustic
line, part of the disk can be mapped onto an arc with colors that do not agree with
that of the counter-image. In some extreme cases, a tangential arc can indeed show no evidence of having been formed by two mirror images (e.g. Smith et al. 2005). Thus a model based only on the analysis of arc colors and their possible associations can lead to results that are not robust. A more detailed study of the color in the arcs, divided into components, is not  the solution and may be wrong because the errors obtained from this kind of measurements become greater as we divide the arc into small fractions. For instance, \citet{alardalone} separated the individual elements in the arcs into two kinds: areas with bright sharp details and areas with smoother images. He found that bright regions have the same colors in the three arcs (as well as the smoother areas), and that the differences in colors between both areas (i.e. comparing the brightest ones with the smoother ones) were statistically significant. Thus, he argued that one single source produces the arc configuration in SL2S\,J02140-0532. However, the magnitudes obtained by performing photometry in regions of 5 pixels in radius could have errors of  $\sim$0.1 mag in bright details, and reach  $\sim$0.3 mag in the smoother areas where the signal is very low. Even with the ACS SnapShots, the errors are substantial, the images are not deep enough for this kind of analysis, and we need to be very cautious about his conclusions. As we discussed above, a model based only on colors would not be robust.  This system clearly shows the need for a spectroscopic redshift to obtain an accurate lens model.

\subsection{Photometric redshifts}

Using the magnitudes reported in Table~\ref{tbl-1}, we estimated the photometric redshifts for the features labeled A, B, and C with the HyperZ  software \citep{hyperz}.  We present the output probability distribution function (PDF) from HyperZ in Fig.~\ref{zphot}. Arc C is constrained to be at $z_{\rm phot} = 1.0 \pm 0.06$, which is in very good agreement with the identification of the emission line as [OII]$\lambda$3727 at $z_{\rm spec}$ = 1.023 $\pm$ 0.001. The multiply imaged system represented by arcs A and B has $z_{\rm phot}$= 1.7 $\pm$ 0.1 and $z_{\rm phot}$= 1.7 $\pm$ 0.1, for the arcs respectively. Although the probability distributions in the three cases are very broad (which is common in photometric redshift estimations using only five filters), there is no overlap in the solutions between system AB and arc C  below a 3$\sigma$ value. Along with the similarity between the distributions of arcs A and B, and the spectroscopic data, the possibility that A, B, and C belong to the same source is again ruled out.

\subsection{Pixelizing the arcs: increasing the number of constraints }\label{Pixelizing}

\begin{figure}[h!]
\begin{center}
\includegraphics[scale=0.6]{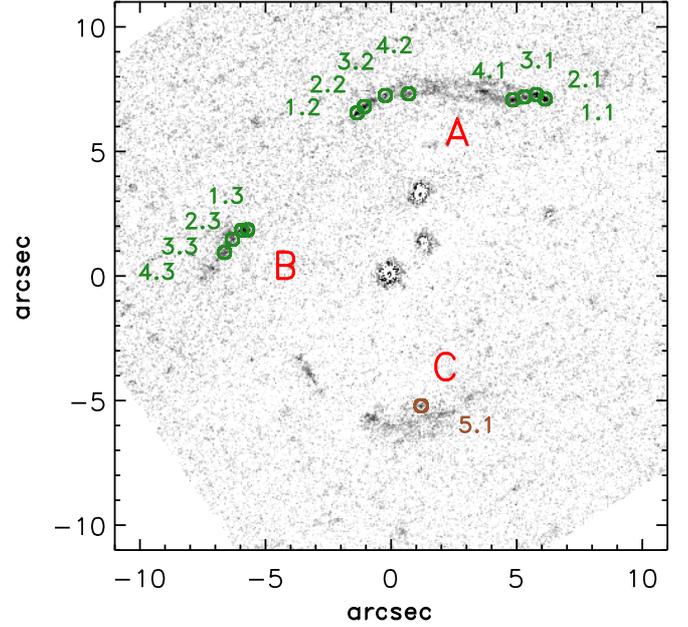}
\caption{F475W image ( 22$\arcsec$$\times$22$\arcsec$) after subtraction of the six central galaxies of the group. The green circles show the positions of the conjugated images that constitutes the AB arc system.  The brown circle, mark the location of arc C.}
\label{fig1b}
\end{center}
\end{figure}

When a source is mapped onto a system that shows multiple subcomponents (surface brightness peaks), similar to the one displayed by \object{SL2S\,J02140-0535} (system AB), the number of constraints can be increased as well as the degrees of freedom (for a fixed number of free parameters), since the peaks can be conjugated as different multiple image systems. 
To highlight the substructures appearing in system AB, we subtracted the  six central galaxies of the group (G1, G2, G3, and the three small ones near those, see Figure~\ref{presentlens} top left). Following \citet{mcl98}, we analyzed the F475W image and fitted a model convolved with a PSF. We used six de Vaucouleurs profiles 
to fit the galaxies and a synthetic PSF because there was no suitable non-saturated star in the field of view. To account for possible differences, we also convolved the PSF with a 
Gaussian profile. In Fig.~\ref{fig1b}, we show the central region of the group after the galaxy subtraction.

To avoid introducing any bias into the modelling, we conjugated any bright peaks for which we were high confidence. This procedure transforms our simple AB system into four different systems, 1.$i$, 2.$i$, 3.$i$, and 4.$i$, where $i$ goes from 1 to 3.
In Fig.~\ref{fig1b}, we show the image identification of the peaks, and in Table~\ref{tbl-2} we report the positions (we assume that the typical uncertainties in the positions of identified conjugated images are 1 pixel $\sim$ 0.05$\arcsec$). To keep the same notation and avoid confusion, we renamed arc C as 5.1. Therefore, our model have five different arc systems in the optimization procedure.

In summary, we have shown that features presented in system AB originate from the 
same background source, which is multiply imaged. This background source at $z_{\rm phot}$ = 1.7 $\pm$ 0.1 yields three images in the image plane: arc A composed of two merging images and arc B. For each image, we have been able to conjugate substructures (systems 1.$i$, 2.$i$, 3.$i$, and 4.$i$). In addition, we have convincingly shown that arc C (now system 5.1) is a different object that is not multiple imaged.

\section{Dynamical constraints}\label{dynamics}

We describe the profiles used in our strong lensing model and also present how the dynamical constraints enter into the model on both the galaxy and group scales. In short, we propose that the dark matter component of the group consists of a single large-scale clump and we add smaller-scale clumps as galaxy-scale perturbations that are associated with the three individual galaxies at the center of \object{SL2S\,J02140-0535}. For these smaller clumps, we  use the internal velocity dispersion of galaxy G1 as well as common scaling laws that links the parameters of the galaxies to their luminosity.  In the large scale-clump, the dynamical information obtained from spectroscopic data is used to set the possible range of values in the parameters that characterize the profile.

\subsection{Galaxy scale}

We assume that the density profile of the stellar mass distribution in the central galaxies of \object{SL2S\,J02140-0535} follows a pseudo isothermal elliptical mass distribution (PIEMD). This profile was derived
by \citet{kassiola} and is parameterized by a central
density $\rho_0$, linked to the central velocity dispersion,
$\sigma_0$, which in turn is related to the depth of the potential
well  \citep[see][for a detailed discussion of the properties of
this mass profile]{mypaperI, ardis2218}.
It is described using two characteristic radii that relate to changes in the
slope of the density profile. In the inner region, the
profile behaves like a non-singular isothermal profile
with central density $\rho_0$ and core radius $r_{core}$.
In the outer parts ($r_{cut} < r$), the density
progressively falls from $\rho \propto r^{-2}$  to
$\rho \propto r^{-4}$, introducing a natural cut-off.
A clump modeled using PIEMD can be
characterized using seven parameters: the center position, ($X,Y$),
the ellipticity $\epsilon$, the position angle
$\theta$, and the parameters of the density profile,
$\sigma_0$, $r_{core}$, and $r_{cut}$. Below, we discuss the parameters associated with the dynamics of the central galaxies, namely, $\sigma_0$ and $r_{cut}$ and we return to those remaining in the next section.

Our spectroscopic data include the two brightest central galaxies in the group, G1 and G2. However, given the poor quality of the G2 galaxy spectra (low S/N ratio), we only measured the line-of-sight internal velocity dispersion of galaxy G1. Using the penalized pixel-fitting  method developed by \citet{Cap04}, we found that $\sigma_{los}^{*}$ = 215 $\pm$34  km s$^{-1}$ for the G-band absorption line profile. Hereafter we denote with an asterisk quantities associated with galaxy G1. This velocity corresponds to $\sigma_{0}^{*}$ = 253 km s$^{-1}$,  according to the relation reported by \citet{ardis2218} who studied the internal velocity dispersion of some galaxies in the Abell\,2218 cluster. 
Assuming that the galaxy profiles can be described by spherically symmetric PIEMD profiles, \citet{ardis2218} found that the ratio of the measured velocity dispersion (which is derived using spectroscopic data) to the fiducial velocity dispersion of the PIEMD profiles, (i.e., the ratio of the stellar to dynamical mass) has an almost  constant value ($\sigma_{0}^{*}$ $\approx$ $\sigma_{los}^{*}$/0.85) across the region where the spectra were obtained.

%%%%%%%%%%%%%%%%%%%%%%%%%%
%%%%%%%%%%%%%%%%%%%%%%%%%%

%%%%%%%%%%%%%%%%%%%%%%%%%%%%%%%%%%%
\begin{table}
\caption{Multiple image identification.\label{tbl-2}}
 % is used to refer this table in the text
\centering % used for centering table
\begin{tabular}{cccc}
\hline\hline % inserts double horizontal lines
\\
Systems &    \multicolumn{1}{c}{$\alpha$}
&  \multicolumn{1}{c}{$\delta$}       \\
&  (J2000)
& (J2000)             \\
\\
\hline % inserts single horizontal line
\\
1.1 & 02:14:07.684  & -05:35:27.37   \\
1.2 & 02:14:08.184 & -05:35:27.54    \\
1.3 & 02:14:08.477 &  -05:35:31.68   \\[3pt]
2.1 & 02:14:07.708  & -05:35:27.19  \\
2.2 & 02:14:08.165 &  -05:35:27.32   \\
2.3 & 02:14:08.493 & -05:35:31.69    \\[3pt]
3.1 & 02:14:07.739 & -05:35:27.26  \\
3.2 & 02:14:08.110 & -05:35:26.95   \\
3.3 & 02:14:08.517 &  -05:35:32.00  \\[3pt]
4.1 & 02:14:07.770 & -05:35:27.35   \\
4.2 & 02:14:08.048 & -05:35:26.93    \\
4.3 & 02:14:08.538 &-05:35:32.48   \\[3pt]
5.1 & 02:14:08.015 & -05:35:38.54  \\
\hline
\end{tabular}
\end{table}
%%%%%%%%%%%%%%%%%%%%%%%%%%%%%%%%%%%

Since the total stellar mass of the galaxy is related to $\sigma_0^{*}$ and $r_{cut}^{*}$ by the expression \citep{ardis2218}

\begin{equation}\label{eq:MassG}
M_{*} = \frac{1.5\pi}{G}r_{cut}^{*}\sigma_0^{*2},
\end{equation}

\noindent the parameter $r_{cut}^{*}$ can be constrained if we determine $M_{*}$ in an independent way. To measure the stellar mass of galaxy G1, we followed two different methods. First, we converted the $g$-band magnitude into an absolute rest frame V-band luminosity and found that $L_V$ = 3.0x10$^{10}$ L$_{\sun V}$. We then used the mass-to-light ratio reported in SLACS lenses \citep{gavazzi07} to estimate the stellar mass and found that $M_{*}$ $\approx$ 1.0x10$^{11}$ M$_{\sun}$. For the second method, we employed the HyperZ  software \citep{hyperz} with the five magnitudes of galaxy G1 reported in Table~\ref{tbl-1}. We found a slightly greater value of $M_{*}$ $\approx$ 4.0x10$^{11}$ M$_{\sun}$. Once we apply Eq.~\ref{eq:MassG}, these range of masses produce a range of possible values for $r_{cut}^{*}$ of 1 - 6 kpc.

The parameters of the two remaining  galaxies are obtained using the scaling relations

\begin{equation}\label{eq:scale}
\begin{array}{l}
 r_{cut}=r_{cut}^{*}(\frac{L}{L^{*}})^{1/2},\\ \\
 \sigma_0 = \sigma_0^{*}(\frac{L}{L^{*}})^{1/4},
\end{array}
\end{equation}

\noindent where $\sigma_0^{*}$, $r_{cut}^{*}$ and $L^{*}$ are set by those values of galaxy G1 discussed above. For a discussion of these scaling relations, we refer to \citet{mypaperIII}.

\subsection{Group scale}

\subsubsection{Velocity dispersion and virial mass}

\begin{figure}[h!]
\begin{center}
\includegraphics[scale=0.45]{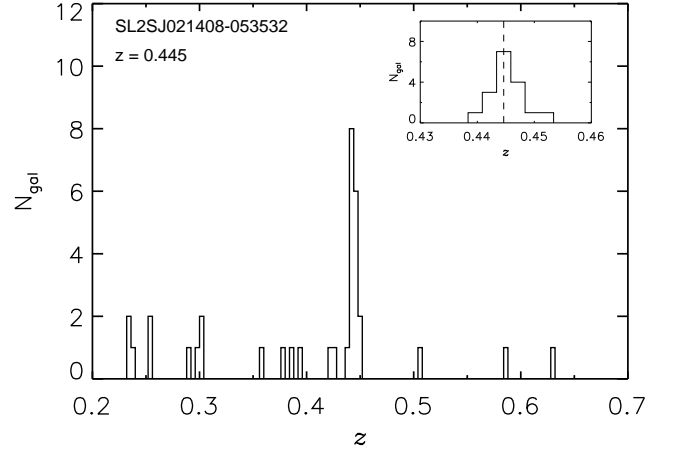}
\caption{The histogram (bin size $\Delta$ = 0.004) shows the redshift distribution along the direction of \object{SL2S\,J02140-0535}. The spectroscopic data reveals a well-defined group at $z$ = 0.444. The inset highlights the central part rebinned to $\Delta$ = 0.0025. There are 17 galaxies in the histogram, but after the iterative algorithm (see text) one outlier was removed.}
\label{z_dist}
\end{center}
\end{figure}

In Fig.~\ref{z_dist}, we show the redshift distribution of the galaxies in the field of \object{SL2S\,J02140-0535}; as we can note, the spectroscopy reveals a well defined group without any evidence of bimodality. The dynamical study of this group will be presented in a forthcoming publication (Mu\~{n}oz et al. 2010 in preparation) as part of a detailed study of seven strong-lensing galaxy groups in the SL2S. Here we briefly describe the method used. We adopted the formalism by \citet{Wil05} in order to determine the group membership. We identified the group members as follows: the group was initially assumed to be located at the redshift of the main bright lens galaxy, $z_{lens}$, with an initial observed-frame velocity dispersion of $\sigma_{obs}$ = 500(1+$z_{lens}$) km\,s$^{-1}$. After computing the required redshift range for group membership and applying a biweight estimator \citep{Bee90} the iterative process was found to have reached a stable membership solution with 16 secure members and $\sigma$ = 630 $\pm$ 107 km\,s$^{-1}$. These galaxies are shown with red rectangles at the bottom of Fig.~\ref{presentlens}.

Assuming that the group has an isotropic velocity dispersion and is no longer undergoing net expansion or contraction, we apply the virial theorem to estimate the mass of \object{SL2S\,J02140-0535}. Since in our lensing models we probe the two-dimensional (2D) mass of the groups, we define $M_v$ = 3$\pi$$\widetilde{R_{v}}$$\sigma_{los}^{2}$/2G  as the virial mass inside the projected radius $\widetilde{R_{v}}$. Using the measured $\sigma_{los}$ = $\sigma/\sqrt{3}$ = 364 km\,s$^{-1}$ and a virial radius $\widetilde{R_{v}}$ = $R_H$ = 0.8 $\pm$ 0.3 Mpc, where $R_H$ is the projected harmonic mean radius \citep[e.g.][]{Irg02}, we obtain a mass of (1.1 $\pm$ 0.8) $\times$ 10$^{14}$ M$_{\sun}$. We stress that this is not the true 2D mass at $\widetilde{R_{v}}$, but only an estimate of the mass inside $\widetilde{R_{v}}$, and that this mass is probably underestimated \citep[e.g.][]{Biv06}.

\subsubsection{Dynamical constraints in the NFW profile}\label{dynamicsNFW}

We describe our galaxy-group density profile as an NFW profile \citep{nav97}

\begin{equation}\label{eq:rho}
\rho(r) = \frac{\rho_{s}}{(r/r_s)(1+r/r_s)^{2}},
\end{equation}

\noindent where $\rho_s$ is a characteristic density, and $r_s$ is a
scale radius that corresponds to the region where the
logarithmic slope of the density equals the isothermal value.
To link the velocity dispersion calculated in the last subsection with the velocity associated with the potential of the NFW halo, we follow two different and independent methods. First we define the velocity in terms of the scale radius and the characteristic density to be

%%%%%%%%%%%%%%%%%%%%%
%%%%%%%%%%%%%%%%%%%%%

\begin{equation}\label{eq:sigmaNFW}
\sigma^{2} \equiv  4\left(1+\ln{\frac{1}{2}}\right)G r_s^{2} \rho_{s}.
\end{equation}

\noindent It is defined in such a way that it
represents a realistic velocity dispersion and
not only a scaling parameter,  which is useful when comparing the velocity predicted by strong lensing models with the velocity dispersion measured in dynamical studies \citep[see the discussion in][]{ver07}. 
In a second way, following \citet{ardis2218} we calculate the projected average line-of-sight velocity dispersion inside the radius $\widetilde{R_{v}}$ as

\begin{equation}\label{eq:sigma2NFW}
\langle \sigma_{los}^{2} \rangle(\widetilde{R_{v}})  = \frac{2\pi \int_{0}^{\widetilde{R_{v}}}\sigma_p^{2}(R)\Sigma(R)RdR}{M_{2D}(\widetilde{R_{v}})},
\end{equation}

\noindent where $R$ is a 2D radius and $\sigma_p^{2}(R)$ is the projected line-of-sight velocity dispersion given by

\begin{equation}\label{eq:sigmaNFW_pro}
 \sigma_{p}^{2} (R)  = \frac{2G}{\Sigma(R)}\int_{R}^{\infty}\frac{M_{3D}(r)\rho(r)}{r^2}\sqrt{r^2-R^2}dr,
\end{equation}

\noindent and $\Sigma(R)$ the surface mass density. We note that Eq.~\ref{eq:sigma2NFW} depends not only on the virial radius, $\widetilde{R_{v}}$, but also in the parameters that characterize the profile. Integrating  Eq.~\ref{eq:rho}, we obtain

\begin{equation}\label{eq:M3D}
 M_{3D}(r)  =   4\pi r_s^{3}\rho_s \left[\ln{(1+r/r_s}) - \frac{r/r_s}{1+r/r_s}\right].
\end{equation}

\noindent Integrating the surface mass density (see Appendix), we then obtain the expression for $M_{2D}(\widetilde{R_{v}})$, and substituting Eqs.~\ref{eq:M3D} and \ref{eq:rho} into Eq.~\ref{eq:sigmaNFW_pro}, it is possible rewrite Eq.~\ref{eq:sigma2NFW} as

\begin{equation}\label{eq:sigma3NFW}
\langle \sigma_{los}^{2} \rangle(\widetilde{R_{v}}) = 2\pi G \rho_s  \frac{G(\widetilde{R_{v}},r_s)}{F(\widetilde{R_{v}},r_s)},
\end{equation}

\noindent where

\begin{eqnarray}\label{eq:GR}
G(\widetilde{R_{v}},r_s) & = & \int_{0}^{\widetilde{R_{v}}}\int_{R}^{\infty}\left[\ln(1+r/r_s)-\frac{r}{r_s+r} \right]  \times
\nonumber\\
& & { } \times  \frac{1}{(r/r_s)(1+r/r_s)^2}\frac{\sqrt{r^2-R^2}}{r^2}R\,dr\,dR
\end{eqnarray}

\noindent and $F(\widetilde{R_{v}},r_s)$ is given by Eq.~\ref{eq:A4}. We emphasize that the dependence of  the projected average line-of-sight velocity dispersion  (Eq.~\ref{eq:sigma3NFW}) on the virial radius, the scale radius, and the characteristic density. If we assume that Eqs.~\ref{eq:sigmaNFW} and \ref{eq:sigma3NFW} are related by the expression $\langle \sigma_{los}^{2} \rangle(\widetilde{R_{v}})$ = $\sigma^2$/3 (which is qualitatively correct in the sense that $\sigma$ represents a 3D velocity dispersion and $\langle \sigma_{los}^{2} \rangle(\widetilde{R_{v}})$ represents the projected average line-of-sight velocity dispersion assuming no anisotropy and a spherically symmetric NFW profile), then

\begin{equation}\label{eq:rv_rs}
1 = \frac{3\pi}{2(1+\ln{1/2})r_s^2}\frac{G(\widetilde{R_{v}},r_s)}{F(\widetilde{R_{v}},r_s)}.
\end{equation}

\noindent  Thus, given  $\widetilde{R_{v}}$, we can use  the above equation to obtain the possible range of values for the scale radius. It is noteworthy that although the velocity dispersion does not appear explicitly in the equation, the equation is not independent of this value since our virial radius is computed from the relative positions of the confirmed members, which in turns depends on the velocity dispersion of the group. In other words, the velocity dispersion is implicit in the equation, and therefore, in our calculations of the scale radius. For  $\widetilde{R_{v}}$ = 0.8 $\pm$ 0.3 Mpc we obtain $r_s$ = 150 $\pm$ 50 kpc. It then follows that $c_{NFW}$ = $\widetilde{R_{v}}$/$r_s$ = 5 $\pm$ 3. After the scale radius is estimated, the 2D mass of the halo expressed in terms of the velocity dispersion of the group and the virial radius can be determined in a simple fashion using Eqs.~\ref{eq:sigmaNFW} and \ref{eq:A3}, and expressed as

\begin{equation}\label{eq:virialmassNFW}
M_{2D}(\widetilde{R_{v}},r_s, \sigma) = \frac{2\pi\sigma^2r_s}{(1+\ln{1/2})G}F(\widetilde{R_{v}},r_s).
\end{equation}

\noindent Using our previously computed values, this  yields $M_{2D}$ = (4 $\pm$ 2) x 10$^{14}$ M$_{\sun}$. This estimate is, within the errors, slightly greater than the value $M_v$ computed from dynamics, although we show in Section~\ref{discuss} that it is in good agreement with the strong and weak lensing mass estimates. For the strong lensing analysis (see Section~\ref{model}), we use these dynamical results to put some priors in the range in which the parameters are allowed to vary: we set the velocity dispersion range to be 523 km\,s$^{-1}$ $\leq$ $\sigma$  $\leq$ 737 km\,s$^{-1}$, and the scale radius to be 100 kpc $\leq$ $r_s$ $\leq$ 200 kpc, i.e. to vary within 1-$\sigma$ of their values.

\section{Gravitational lensing analysis}\label{model}

\subsection{Strong lensing}

To study the lensing group \object{SL2S\,J02140-0535}, we used
the parametric method implemented in the LENSTOOL\footnote{
This software is publicly available at:
http://www.oamp.fr/cosmology/ lenstool/} ray-tracing code
\citep{jphd}. This software use a Bayesian Monte Carlo Markov chain (MCMC) method to search for the
most likely parameters in the lens modeling \citep{jullo07}.

%%%%%%%%%%%%%%%%%%%%%
\begin{figure}[h!]
\begin{center}
\includegraphics[scale=0.5]{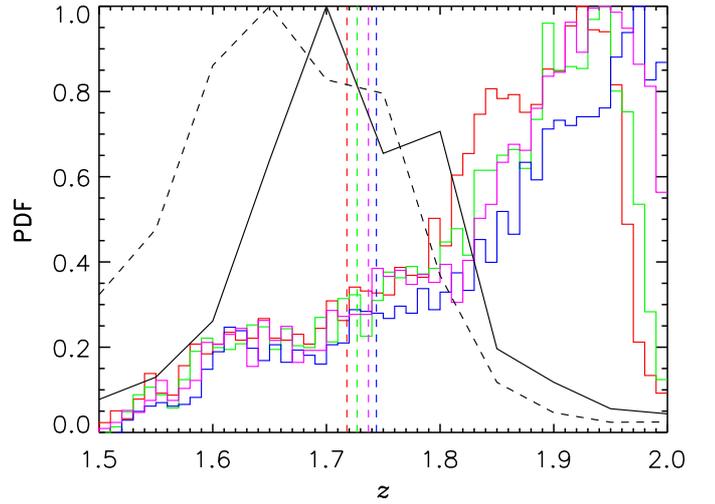}
\caption{Normalized PDFs. The red, blue, green and magenta histograms shows the distributions obtained with our model, for $z_1$, $z_2$, $z_3$ and $z_4$ respectively. The vertical lines corresponds to the best-fit model solutions given by the optimization procedure. Black continuous and black dashed lines are the distributions shown in Figure~\ref{zphot} for arcs A and B, respectively.}
\label{arcz}
\end{center}\end{figure}
%%%%%%%%%%%%%%%%%%%%%

%%%%%%%%%%%%%%%%%%%%%%%%%%
\begin{figure}[h!]
\begin{center}
\includegraphics[scale=0.6]{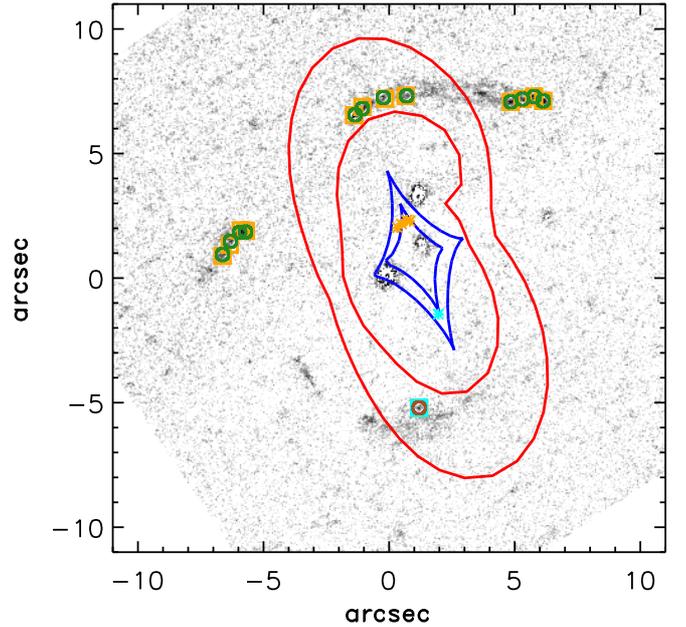}
\caption{Tangential critical lines (red) and caustic lines (blue) for two different sources located at  $z$ = 1.0 and $z$ = 1.7. As in Fig.~\ref{fig1b}, the circles show the measured positions of the images (input data for the model). The squares (orange and cyan) are the model-predicted image positions after optimization, and the asterisks (orange and cyan) the positions in the source plane. Although the cyan asterisk appears to over lay the caustic line, this is only an artifact, a product of the thick lines, and the size of the asterisks that we used to highlight the figure. The source is \textit{not} over the caustic line.}
\label{Modelo1}
\end{center}
\end{figure}
%%%%%%%%%%%%%%%%%%%%%%%%%%

%%%%%%%%%%%%%%%%%%%%%%%%%%%%%%%%%%   

%\begin{landscape}
\begin{table*}
\caption{Best-fit model parameters.}
\label{tbl-3} % is used to refer this table in the text
\centering % used for centering table
\begin{tabular}{lcccccccccccc}
\hline\hline % inserts double horizontal lines
\\

 Comp.   &  \multicolumn{1}{c}{X} &
\multicolumn{1}{c}{Y}  &   \multicolumn{1}{c}{$\epsilon$}   &
\multicolumn{1}{c}{$\theta$} & \multicolumn{1}{c}{$r_{core}$}  &  \multicolumn{1}{c}{$r_s$}  &
\multicolumn{1}{c}{$r_{cut}$} & \multicolumn{1}{c}{$c_{200}$} & \multicolumn{1}{c}{$\sigma_0$}
 & \multicolumn{1}{c}{$\sigma$}
& $\chi^{2}_{DOF}$  
\\

  &  \multicolumn{1}{c}{[\arcsec]} &
\multicolumn{1}{c}{[\arcsec]}  &        &
\multicolumn{1}{c}{[$^{\circ}$]} & \multicolumn{1}{c}{[kpc]}  &  \multicolumn{1}{c}{[kpc]}  &
 \multicolumn{1}{c}{[kpc]} & &  \multicolumn{1}{c}{[km s$^{-1}$]}
& \multicolumn{1}{c}{[km s$^{-1}$]}  
&

\\
\hline % inserts single horizontal line
\\

 $$ & $$                        & $$                & $$                                    & $$                                               & $$    & $$  & $$    & $$    &  $$   &  & $1.2$           \\
    Group  & $1.3\pm0.1$ & $0.5\pm0.3$ & $0.28\pm0.02$  & $111.6\pm0.2$    & ---  & $170\pm18$    & --- & $6.0\pm0.6$   & --- & $664\pm18$  &  $$      \\
\\

   $L^{*}$  & ---              & ---                     & ---        & ---                              & $[0.15]$       & ---    & $5.6\pm1.3$                           &         & $[253]$         & ---      &  $$               \\

\\

\\
\\
\hline %inserts single line
\end{tabular}
\tablefoot{
The first column identifies the different scale components. The $L^{*}$ denotes the galaxy-scale mass component. Columns 2 and 3 list the position in
arcseconds relative to the BGG. In columns 4 and 5 we provide the geometric parameters.
From columns 6 to 11, we present the different profiles parameters, and in the last column
the  $\chi^{2}_{DOF}$. Values in square brackets are not optimized.
}
\end{table*}
%\end{landscape}

 %%%%%%%%%%%%%%%%%%%%%%%%%%%%%%%%        

The identification of the multiply imaged system presented in Section\,\ref{data}  leads to 16 observational constraints. With these constraints, we computed a model (optimized in the image plane) with the following set of free parameters \{$X$,  $Y$, $\epsilon$,  $\theta$, $r_s$, $\sigma$,  $z_1$,  $z_2$, $z_3$, $z_4$, $r_{cut}^{*}$\}, where the first six parameters characterize the NFW profile and $z_i$ represents the redshifts for the systems 1.$i$, 2.$i$, 3.$i$, and 4.$i$, respectively. All the parameters are allowed to vary with uniform priors except $r_s$ and $\sigma$ for which we used Gaussian priors with the 1-$\sigma$ errors computed in the last section. System 5.1, at $z_{\rm spec}$ = 1.023 $\pm$ 0.001 is used in the analysis as a singly imaged object. Multiple image systems were allowed to vary between 1.5 and 2.0 (approximately 2-$\sigma$ from the value inferred from the photometric redshift analysis) to take into account the broad PDF common to these measurements; and for the same reason we do not choose a Gaussian prior in this case. The remaining parameters, those that describe the central galaxies, were set as follows: the center of the profiles and the ellipticity and position angle are assumed to be the same as for the luminous components. We use the values reported in Table~\ref{tbl-1}. The velocities dispersions are given by Eq.~\ref{eq:scale} for a $\sigma_0^{*}$ = 253 km\,s$^{-1}$.

Figure~\ref{arcz}  shows the PDF for the arc systems, and the best-fit model solutions for $z_i$ obtained after the optimization. The mean optimal value is $\bar{z}$ = 1.7 $\pm$ 0.2. It is clear from the figure that the model PDFs are slightly far to be Gaussians and $z$ values greater than two are still favored. This reflects the lack of arc systems (only one system with a fixed redshift) available to constrain the model and shows the importance of using the photometric redshift as a prior. The results of our best fits are summarized in Table~\ref{tbl-3}. In Figure~\ref{Modelo1}, we show the predicted positions of our best model (orange and cyan squares), as well as the observed positions (green and brown circles). The remarkable agreement between both positions is quantified by a mean scatter in the image plane smaller than 0.04$\arcsec$. We obtained $\chi^2_{DOF}$ = 1.2, showing that the extent of the match between the observed positions and those predicted by the model is in accord with the error variance. This $\chi^2_{DOF}$ is defined in the image plane \citep[see][]{jullo07} assuming the same uncertainties in the positions for all the images (see Sect.\,\ref{Pixelizing}). We note that our model do not predict extra images, which supports the reliability of our strong lensing model.

\subsection{Weak lensing}
The weak lensing analysis of this galaxy group was performed in \citet{paperI}.
We refer the interested reader to that paper for a detailed description of the
methodology, and present below a brief summary.
They selected as background sources the galaxies whose $i$ band magnitude falls between 21.5 and 24.
Their density equals to 13 arcmin$^{-2}$.
The completeness magnitude in this band is 23.96 and the seeing equals  0.61$\arcsec$. 
A Bayesian method, implemented in the \textsc{Im2shape} software \citep{im2shape}, is used to fit the shape
parameters of the faint background galaxies and correct for PSF smearing.
From the catalogue of background galaxies, \citet{paperI} pursued a one-dimensional weak lensing analysis.
They fitted a singular isothermal sphere model (SIS) to the reduced shear signal between
150\,kpc and 1.2\,Mpc from the group center,
finding an Einstein radius equals to 3.6 $\pm$ 2.4$\arcsec$.
To relate the strength of the weak lensing signal to a physical velocity dispersion characterising
the group potential, \citet{paperI} estimated the mean geometrical factor using the photometric 
redshift catalogue from the T0004 release of the CFHTLS-Deep survey\footnote{http://www.ast.obs-mip.fr/users/roser/CFHTLS\_T0004/} \citep{iena}.
They found that $\sigma_{\rm SIS} = 612^{+180}_{-264}$ km\,s$^{-1}$.
This translates into a projected mass computed within a circular aperture of radius equals to 1\,Mpc of (2.8$ \pm $1.5) $\times$ 10$^{14}$ M$_{\sun}$.

\section{Discussion}\label{discuss}

Our best-fit model demonstrates that it is possible to recover the observational features presented in the galaxy group \object{SL2S\,J02140-0535} using a single halo for the group component plus three galaxy scale components associated with the central galaxies. We do not need extra assumptions about the mass distribution in the group. In addition, the position angle of the halo is oriented with a well-constrained position angle  at $\theta$ = 111.6$\pm$0.2, which is the same direction as the one defined by the luminosity contours \citep{paperI} and is consistent with the spatial distribution of the confirmed members of the group (see Fig.~\ref{presentlens} bottom). This supports a scenario where the mass is traced by light.

\begin{figure}[h!]
\begin{center}
\includegraphics[scale=0.6]{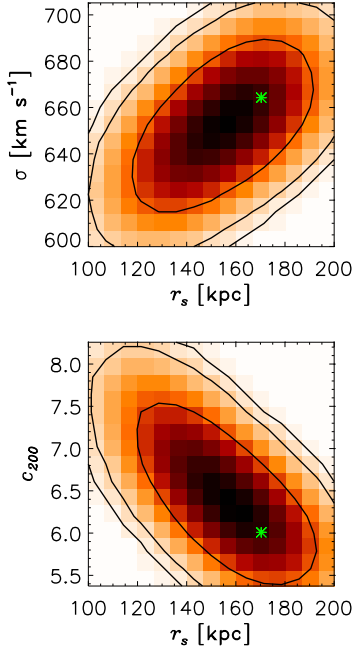}
\caption{PDFs of the parameters $\sigma$, $c_{200}$ and $r_s$. The three contours stand for the 68$\%$, 95$\%$, and 99$\%$ confidence levels. The values obtained for our best-fit model are marked by a green asterisk.}
\label{degeneracy}
\end{center}\end{figure}

The degeneracies in the $\sigma$-$r_s$ space are very common in strong lensing modeling \citep[e.g.][]{jullo07} since the region where the arcs appear is limited to the central region of the clusters and the scale radius is generally three or four times the distance to that region. In galaxy groups, this region is indeed, even smaller than in clusters. Thus, it is impossible to constrain the scale radius of a group using only its observed strong lensing features. To highlight this behavior, we depicted in the top panel of Figure~\ref{degeneracy} the PDFs for the $\sigma$ and $r_s$ parameters. We note how models with a scale radius greater than 200 kpc are still possible (in addition to those smaller than 100 kpc), but we placed a limit on this value using information from the group dynamics. Thus, this give us a method that would be useful in breaking the degeneracies in the $\sigma$-$r_s$ space using the dynamics of the group obtained from spectroscopic data. This is important because it provides a technique to fill the gap between strong lensing and weak lensing constraints in galaxy groups.

%%%%%%%%%%%%%%%%%%%%%
%%%%%%%%%%%%%%%%%%%%%

%%%%%%%%%%%%%%%%%%%%%
\begin{figure}[h!]
\begin{center}
\includegraphics[scale=0.5]{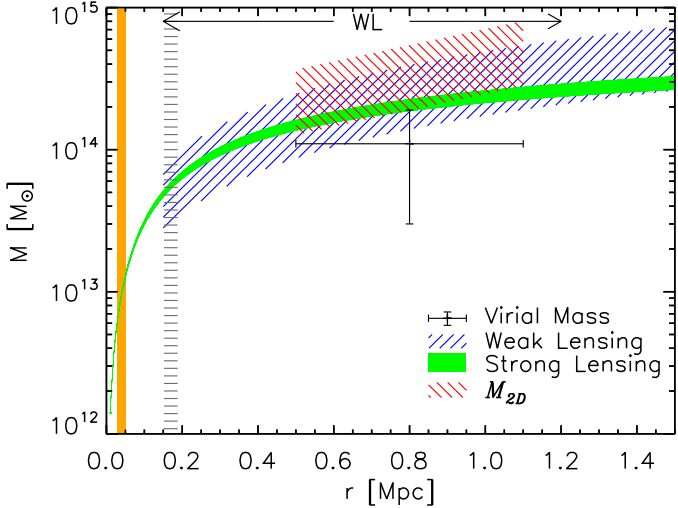}
\caption{The 2D projected mass as a function of the aperture radius
measured from the BGG. The green and blue shaded areas corresponds to the mass profile within 1$\sigma$ errors for the strong lensing and the weak lensing models, respectively. The red region  illustrates the 2D mass derived from Eq.~\ref{eq:virialmassNFW} as a function of $\widetilde{R_{v}}$. The point with error bars is the mass estimate from dynamics. The orange-shaded region shows the area where the arc systems  lie and the gray region shows the model-predicted scale radius.  We also depict with an arrow the zone in which the weak lensing signal was measured.}
\label{MvsR}
\end{center}\end{figure}
%%%%%%%%%%%%%%%%%%%%%

To analyze the dark matter profile of  \object{SL2S\,J02140-0535} in greater detail, we calculated the concentration $c_{200}$ from the values of $r_s$ and $\sigma$ \citep[see the discussion in][]{ver07}. Here we assumed that  $c_{200}$ = $r_{200}/r_s$, where we defined $r_{200}$ as the radius of a spherical volume within which the mean density is 200 times the critical density of the universe at the given redshift $z$ of the group. The bottom panel of Figure~\ref{degeneracy} shows the 2D histogram for the parameters $c_{200}$ and $r_s$, where we can note the same tendency with $r_s$ as in the top panel. Our best-fit model infers that $c_{200}$ = 6.0 $\pm$ 0.6, which is in good agreement with the value $c_{NFW}$ presented in Section~\ref{model}. On the other hand, if we compare it with the values expected from $\Lambda$CDM numerical simulations, we found that our concentration is slightly higher.   From the work of \citet{Duff08}, a DM halo at $z$ = 0.44 with $M_{200}$ $\approx$ 1 $\times$ 10$^{14}$ M$_{\sun}$ has $c_{200}$  $\approx$ 4.0.

Given the relatively low mass of galaxy groups (compared with clusters), we do not expect to derive a weak lensing signal strong enough to allow us to fully probe an NFW profile. Until the work of \citet{paperI}, weak lensing signals on these mass scales had indeed been recovered only by stacking group images \citep[e.g.][]{rachelgroup}. However, the strength of the signal in \object{SL2S\,J02140-0535} is sufficient to fit a SIS and obtain the projected mass as a function of the radius. In Figure~\ref{MvsR},  we compare the masses obtained with the different methods adopted in the present work:  strong lensing (green), weak lensing (blue), $M_{2D}$ derived from Eq.~\ref{eq:virialmassNFW} (red), and the virial mass (point with error bars). As we place limits on the possible values of the scale radius, the strong lensing mass is still reliable up to the scale radius ($r_s$ = 170 $\pm$ 18 kpc), but beyond this value the extrapolation of the mass becomes less secure. However, from Figure~\ref{MvsR} we can note that the first three masses are in agreement within the errors in the range 0.5 Mpc $\leq$ $r$ $\leq$ 1.1 Mpc, i.e. the interval defined by  $\widetilde{R_{v}}$. The virial mass, despite being lower than all masses, differs more from $M_{2D}$. However, we stress again that it is not the true 2D mass at $\widetilde{R_{v}}$ but only an estimate of the mass inside $\widetilde{R_{v}}$. This mismatch between virial mass and lensing mass is also found in other groups, but the analysis of this subject is beyond the scope of the present work and  will be discussed in detail in a forthcoming paper (Mu\~{n}oz et al. 2010 in preparation).

Using a perturbative method, \citet{alardalone}
reconstructed the structure and potential of the gravitational lens \object{SL2S\,J02140-0535}. In his work, he assumed that the A, B, and C arcs belong to the same source. He argued that the differences in colors observed in the HST-ACS images were related to the crossing of the caustic line by part of the source. His model predicts that one half of image C comes from the same part of the source as images A and B, but the other half corresponds to an area of the source that has no counterpart in the large arcs (A and B). The perturbative method is capable of reproducing this arc configuration, but the model predicts the existence of 
an independent dark component that does not follow light.  In the present work, we have demonstrated that AB system and C system belong to different sources, one at $z$ = 1.7 $\pm$ 0.1 and the other one at $z_{spec}$ = 1.023 $\pm$ 0.001. Our model, with a single NFW profile (which is consistent with the analysis of the luminosity contours, Fig.~\ref{presentlens} bottom, and the redshift distribution of the galaxies in the field of \object{SL2S\,J02140-0535}, see Fig.~\ref{z_dist}), provide a good fit of the arc features observed in the group and our results point out that the projected light of the group traces the dark matter distribution.

\section{Conclusions}\label{conclus}

We have probed the gravitational potential of \object{SL2S\,J02140-0535} on a large
radial range using complementary techniques: strong lensing, weak lensing, and dynamics.
In the strong lensing regime, we performed a fit using an NFW profile and
three galaxy-scale mass components as perturbations to the group potential. We used
PIEMDs for the individual galaxies and set constraints on their geometrical and dynamical parameters using observational data (image and spectroscopy). We measured the velocity dispersion of the group using Keck (LRIS) and VLT (FORS2) spectra,  and we applied this information to constrain and  build a reliable strong lensing model based on one multiply imaged system and one single image system. In the weak lensing regime, we used the analysis reported in \citet{paperI}. Our main results can be summarized as follows:

\begin{enumerate}

   \item We have shown that the AB and C systems belong to different sources, one at $z$ = 1.7 $\pm$ 0.1 (photometric redshift) and the other one at $z$ = 1.023 $\pm$ 0.001 (spectroscopic), respectively.

   \item Our best-fit lens model reproduce quite well the image systems observed in the field of \object{SL2S\,J02140-0535}, with no extra assumptions about the underlying mass distribution (e.g. bimodal distribution).
This agrees with the spectroscopic analysis that shows a well-defined group. Furthermore, we obtained a position angle of the halo of $\theta=111.6\pm0.2$, that is consistent with the orientation of the luminosity contours and the spatial distribution of confirmed members.

  \item The spectroscopic information provide constraints on the scale radius of the NFW profile. In the present work, we showed that is possible to use the dynamics at group scale to discriminate between a great range of possible $r_s$ values and $\sigma$ (in particular  100 kpc $\leq$ $r_s$ $\leq$ 200 kpc and 523 km\,s$^{-1}$ $\leq$ $\sigma$  $\leq$ 737 km\,s$^{-1}$, respectively).

   \item The scale radius and the velocity dispersion found in our best-fit models, $r_s$ = 170 $\pm$ 18 kpc and $\sigma$ = 664 $\pm$ 18 km\,s$^{-1}$, respectively, yields a  concentration value $c_{200}$ = 6.0 $\pm$ 0.6, which is consistent with $c_{NFW}$ = 5 $\pm$ 3, but slightly greater that the one predicted by $\Lambda$CDM simulations.

   \item  The masses obtained with the strong lensing, weak lensing, and $M_{2D}$ methods agree within the errors, but the virial mass is slightly below these values. Thus, we demonstrated that it is possible to combine these  different methodologies to get a complete insight into the mass of a galaxy group.

\end{enumerate}

The lensing and dynamical analysis of  \object{SL2S\,J02140-0535} presented in this
paper,  has had two principal points. First, it has investigated the impact of including spectroscopic
information of arcs (i.e. secure arc identifications) in strong lensing models to avoid complex mass distribution models 
that do not follow the light. Secondly, we studied the possibility of constraining large scale properties that could not be measured by strong lensing, such as the scale radius in the NFW profile.
Future spectroscopic follow-up of the arcs A and B
will provide confirmation of our model. In addition, the deepest HST images of the group will allow us to construct more accurate weak lensing models capable of probing an NFW profile on large scales, as well as  improving our strong lensing model.

\begin{acknowledgements}
We thank the anonymous referee for thoughtful comments and suggestions.    T. Verdugo acknowledge support  from FONDECYT through grant 3090025. T. Verdugo thank the Laboratoire d'Astrophysique de Marseille and the Institut d'Astrophysique de Paris for the two kindly invitations to work in their facilities. VM acknowledges partial support from FONDECYT 1090673 and DIPUV 09/2007. RPM acknowledges partial support from CONICYT CATA-BASAL and Comit\'e Mixto ESO-Gobierno de Chile.
ML thanks the center National d'Etudes Spatiales (CNES) and CNRS for their support. ML est b\'en\'eficiaire d'une bourse d'acceuil de la Ville de Marseille. The Dark Cosmology center is funded by the Danish National National Research Foundation. JR acknowledges support from a EU Marie-Curie fellowship. We also thank R. Pell{\' o} for helping with HyperZ.
\end{acknowledgements}

\begin{appendix}\label{appendix} 
\section{M$_{2D}$ inside the virial radius $\widetilde{R_{v}}$ in the NFW profile}

The surface mass density in the NFW profile is given by \citep{gol02}

\begin{equation}\label{eq:A1}
\Sigma(\xi) = 2\rho_sr_sF(\xi),
\end{equation}

\noindent where

\begin{equation}\label{eq:A2}
F(\xi) = \left\{ \begin{array}{ll}
\frac{1}{\xi^2-1}   \left( 1 - \frac{1}{\sqrt{1-\xi^{2}}} \, \rm{arccosh}\frac{1}{\xi} \right) & \textrm{if $\xi<1$}\\

\frac{1}{3} & \textrm{if $\xi = 1$}\\

\frac{1}{\xi^2-1}   \left(1 - \frac{1}{\sqrt{ \xi^{2}-1}}\arccos\frac{1}{\xi} \right) & \textrm{if $\xi>1$}\\
\end{array} \right.
\end{equation}

\noindent  and the dimensionless coordinate $\xi$, is the radius in the $XY$ plane in units of the scale radius,  $\xi$ = ($x/r_s$, $y/r_s$). It then follows that the 2D mass inside the virial radius can be expressed as:

\begin{equation}\label{eq:A3}
M_{2D}(\rho_s,\widetilde{R_{v}},r_s) = 8 \pi \rho_s r_s^3F(\widetilde{R_{v}},r_s),
\end{equation}

\noindent where

\begin{eqnarray}\label{eq:A4}
F(\widetilde{R_{v}},r_s) & = & \int_{0}^{1} \frac{1}{\sqrt{\xi^2-1}}\left[1- \frac{1}{\sqrt{1-\xi^2}} \, \rm{arccosh}\frac{1}{\xi} \right]\,d\xi +
\nonumber\\
& &  \int_{1}^{\widetilde{\xi}}\frac{1}{\sqrt{\xi^2-1}}\left[1- \frac{1}{\sqrt{\xi^2-1}}\arccos\frac{1}{\xi}  \right]\,d\xi.
\end{eqnarray}

\noindent and $\widetilde{\xi}$ = $\widetilde{R_{v}}/r_s$. We note that the mass depends on both the characteristic density and the scale radius of the NFW profile.

\end{appendix}

\bibliographystyle{aa} % style aa.bst
%\bibliography{references} % your references Yourfile.bib
\bibliography{references}

\begin{thebibliography}{46}
\expandafter\ifx\csname natexlab\endcsname\relax\def\natexlab#1{#1}\fi

\bibitem[{{Alard}(2009)}]{alardalone}
{Alard}, C. 2009, \aap, 506, 609

\bibitem[{{Beers} {et~al.}(1990){Beers}, {Flynn}, \& {Gebhardt}}]{Bee90}
{Beers}, T.~C., {Flynn}, K., \& {Gebhardt}, K. 1990, \aj, 100, 32

\bibitem[{{Bertin} \& {Arnouts}(1996)}]{sextractor}
{Bertin}, E. \& {Arnouts}, S. 1996, \aap, 117, 393

\bibitem[{{Biviano} {et~al.}(2006){Biviano}, {Murante}, {Borgani}, {Diaferio},
  {Dolag}, \& {Girardi}}]{Biv06}
{Biviano}, A., {Murante}, G., {Borgani}, S., {et~al.} 2006, \aap, 456, 23

\bibitem[{{Bolzonella} {et~al.}(2000){Bolzonella}, {Miralles}, \& {Pell{\'
  o}}}]{hyperz}
{Bolzonella}, M., {Miralles}, J.-M., \& {Pell{\' o}}, R. 2000, \aap, 363, 476

\bibitem[{{Bridle} {et~al.}(2002){Bridle}, {Kneib}, {Bardeau}, \&
  {Gull}}]{im2shape}
{Bridle}, S., {Kneib}, J.-P., {Bardeau}, S., \& {Gull}, S. 2002, in The shapes
  of galaxies and their dark halos, Proceedings of the Yale Cosmology Workshop
  , New Haven, Connecticut, USA, 28-30 May 2001. Edited by Priyamvada
  Natarajan., ed. P.~{Natarajan}, 38--+

\bibitem[{{Cabanac} {et~al.}(2007){Cabanac}, {Alard}, {Dantel-Fort}, {Fort},
  {Gavazzi}, {Gomez}, {Kneib}, {Le F{\`e}vre}, {Mellier}, {Pello}, {Soucail},
  {Sygnet}, \& {Valls-Gabaud}}]{sl2s}
{Cabanac}, R.~A., {Alard}, C., {Dantel-Fort}, M., {et~al.} 2007, \aap, 461, 813

\bibitem[{{Cappellari} \& {Emsellem}(2004)}]{Cap04}
{Cappellari}, M. \& {Emsellem}, E. 2004, \pasp, 116, 138

\bibitem[{{D'Onghia} {et~al.}(2005){D'Onghia}, {Sommer-Larsen}, {Romeo},
  {Burkert}, {Pedersen}, {Portinari}, \& {Rasmussen}}]{elena05}
{D'Onghia}, E., {Sommer-Larsen}, J., {Romeo}, A.~D., {et~al.} 2005, \apjl, 630,
  L109

\bibitem[{{Duffy} {et~al.}(2008){Duffy}, {Schaye}, {Kay}, \& {Dalla
  Vecchia}}]{Duff08}
{Duffy}, A.~R., {Schaye}, J., {Kay}, S.~T., \& {Dalla Vecchia}, C. 2008,
  \mnras, 390, L64

\bibitem[{{El{\'{\i}}asd{\'o}ttir} {et~al.}(2007){El{\'{\i}}asd{\'o}ttir},
  {Limousin}, {Richard}, {Hjorth}, {Kneib}, {Natarajan}, {Pedersen}, {Jullo},
  \& {Paraficz}}]{ardis2218}
{El{\'{\i}}asd{\'o}ttir}, {\'A}., {Limousin}, M., {Richard}, J., {et~al.} 2007,
  ArXiv e-prints, 710

\bibitem[{{Faltenbacher} \& {Mathews}(2007)}]{group7}
{Faltenbacher}, A. \& {Mathews}, W.~G. 2007, \mnras, 375, 313

\bibitem[{{Finoguenov} {et~al.}(2007){Finoguenov}, {Ponman}, {Osmond}, \&
  {Zimer}}]{group5}
{Finoguenov}, A., {Ponman}, T.~J., {Osmond}, J.~P.~F., \& {Zimer}, M. 2007,
  \mnras, 374, 737

\bibitem[{{Gastaldello} {et~al.}(2007){Gastaldello}, {Buote}, {Humphrey},
  {Zappacosta}, {Bullock}, {Brighenti}, \& {Mathews}}]{fabio}
{Gastaldello}, F., {Buote}, D.~A., {Humphrey}, P.~J., {et~al.} 2007, \apj, 669,
  158

\bibitem[{{Gavazzi} {et~al.}(2007){Gavazzi}, {Treu}, {Rhodes}, {Koopmans},
  {Bolton}, {Burles}, {Massey}, \& {Moustakas}}]{gavazzi07}
{Gavazzi}, R., {Treu}, T., {Rhodes}, J.~D., {et~al.} 2007, \apj, 667, 176

\bibitem[{{Golse} \& {Kneib}(2002)}]{gol02}
{Golse}, G. \& {Kneib}, J. 2002, \aap, 390, 821

\bibitem[{{Helsdon} \& {Ponman}(2000)}]{group1}
{Helsdon}, S.~F. \& {Ponman}, T.~J. 2000, \mnras, 315, 356

\bibitem[{{Helsdon} \& {Ponman}(2003)}]{group2}
{Helsdon}, S.~F. \& {Ponman}, T.~J. 2003, \mnras, 339, L29

\bibitem[{{Ienna} \& {Pell{\'o}}(2006)}]{iena}
{Ienna}, F. \& {Pell{\'o}}, R. 2006, in SF2A-2006: Semaine de l'Astrophysique
  Francaise, ed. D.~{Barret}, F.~{Casoli}, G.~{Lagache}, A.~{Lecavelier}, \&
  L.~{Pagani}, 347--+

\bibitem[{{Irgens} {et~al.}(2002){Irgens}, {Lilje}, {Dahle}, \&
  {Maddox}}]{Irg02}
{Irgens}, R.~J., {Lilje}, P.~B., {Dahle}, H., \& {Maddox}, S.~J. 2002, \apj,
  579, 227

\bibitem[{{Jullo} {et~al.}(2007){Jullo}, {Kneib}, {Limousin},
  {El{\'{\i}}asd{\'o}ttir}, {Marshall}, \& {Verdugo}}]{jullo07}
{Jullo}, E., {Kneib}, J.-P., {Limousin}, M., {et~al.} 2007, New Journal of
  Physics, 9, 447

\bibitem[{{Kassiola} \& {Kovner}(1993)}]{kassiola}
{Kassiola}, A. \& {Kovner}, I. 1993, \apj, 417, 450

\bibitem[{{Kneib}(1993)}]{jphd}
{Kneib}, J.-P. 1993, PhD thesis, Universit\'e Paul Sabatier, Toulouse III,
  France

\bibitem[{{Limousin} {et~al.}(2009){Limousin}, {Cabanac}, {Gavazzi}, {Kneib},
  {Motta}, {Richard}, {Thanjavur}, {Foex}, {Pello}, {Crampton}, {Faure},
  {Fort}, {Jullo}, {Marshall}, {Mellier}, {More}, {Soucail}, {Suyu},
  {Swinbank}, {Sygnet}, {Tu}, {Valls-Gabaud}, {Verdugo}, \& {Willis}}]{paperI}
{Limousin}, M., {Cabanac}, R., {Gavazzi}, R., {et~al.} 2009, \aap, 502, 445

\bibitem[{{Limousin} {et~al.}(2010){Limousin}, {Jullo}, {Richard}, {Cabanac},
  {Suyu}, {Halkola}, {Kneib}, {Gavazzi}, \& {Soucail}}]{my0854}
{Limousin}, M., {Jullo}, E., {Richard}, J., {et~al.} 2010, \aap, 524, A95+

\bibitem[{{Limousin} {et~al.}(2005){Limousin}, {Kneib}, \&
  {Natarajan}}]{mypaperI}
{Limousin}, M., {Kneib}, J.-P., \& {Natarajan}, P. 2005, \mnras, 356, 309

\bibitem[{{Limousin} {et~al.}(2007){Limousin}, {Richard}, {Jullo}, {Kneib},
  {Fort}, {Soucail}, {El{\'{\i}}asd{\'o}ttir}, {Natarajan}, {Ellis}, {Smail},
  {Czoske}, {Smith}, {Hudelot}, {Bardeau}, {Ebeling}, {Egami}, \&
  {Knudsen}}]{mypaperIII}
{Limousin}, M., {Richard}, J., {Jullo}, E., {et~al.} 2007, \apj, 668, 643

\bibitem[{{Mamon}(2007)}]{mamongroup}
{Mamon}, G.~A. 2007, in Groups of Galaxies in the Nearby Universe, ed.
  I.~{Saviane}, V.~D. {Ivanov}, \& J.~{Borissova}, 203--+

\bibitem[{{Mandelbaum} {et~al.}(2006){Mandelbaum}, {Seljak}, {Cool}, {Blanton},
  {Hirata}, \& {Brinkmann}}]{rachelgroup}
{Mandelbaum}, R., {Seljak}, U., {Cool}, R.~J., {et~al.} 2006, \mnras, 372, 758

\bibitem[{{McKean} {et~al.}(2010){McKean}, {Auger}, {Koopmans}, {Vegetti},
  {Czoske}, {Fassnacht}, {Treu}, {More}, \& {Kocevski}}]{mck10}
{McKean}, J.~P., {Auger}, M.~W., {Koopmans}, L.~V.~E., {et~al.} 2010, \mnras,
  317

\bibitem[{{McLeod} {et~al.}(1998){McLeod}, {Bernstein}, {Rieke}, \&
  {Weedman}}]{mcl98}
{McLeod}, B.~A., {Bernstein}, G.~M., {Rieke}, M.~J., \& {Weedman}, D.~W. 1998,
  \aj, 115, 1377

\bibitem[{{Navarro} {et~al.}(1997){Navarro}, {Frenk}, \& {White}}]{nav97}
{Navarro}, J.~F., {Frenk}, C.~S., \& {White}, S.~D.~M. 1997, \apj, 490, 493

\bibitem[{{Newman} {et~al.}(2009){Newman}, {Treu}, {Ellis}, {Sand}, {Richard},
  {Marshall}, {Capak}, \& {Miyazaki}}]{new09}
{Newman}, A.~B., {Treu}, T., {Ellis}, R.~S., {et~al.} 2009, \apj, 706, 1078

\bibitem[{{Oke} {et~al.}(1995){Oke}, {Cohen}, {Carr}, {Cromer}, {Dingizian},
  {Harris}, {Labrecque}, {Lucinio}, {Schaal}, {Epps}, \& {Miller}}]{lris}
{Oke}, J.~B., {Cohen}, J.~G., {Carr}, M., {et~al.} 1995, \pasp, 107, 375

\bibitem[{{Osmond} \& {Ponman}(2004)}]{group3}
{Osmond}, J.~P.~F. \& {Ponman}, T.~J. 2004, \mnras, 350, 1511

\bibitem[{{Rasmussen} \& {Ponman}(2007)}]{group6}
{Rasmussen}, J. \& {Ponman}, T.~J. 2007, \mnras, 380, 1554

\bibitem[{{Sand} {et~al.}(2002){Sand}, {Treu}, \& {Ellis}}]{sand02}
{Sand}, D.~J., {Treu}, T., \& {Ellis}, R.~S. 2002, \apjl, 574, L129

\bibitem[{{Sand} {et~al.}(2004){Sand}, {Treu}, {Smith}, \& {Ellis}}]{sand04}
{Sand}, D.~J., {Treu}, T., {Smith}, G.~P., \& {Ellis}, R.~S. 2004, \apj, 604,
  88

\bibitem[{{Sommer-Larsen}(2006)}]{jespergroup}
{Sommer-Larsen}, J. 2006, \mnras, 369, 958

\bibitem[{{Thanjavur} {et~al.}(2010){Thanjavur}, {Crampton}, \&
  {Willis}}]{tha10}
{Thanjavur}, K., {Crampton}, D., \& {Willis}, J. 2010, ArXiv e-prints

\bibitem[{{Tu} {et~al.}(2009){Tu}, {Gavazzi}, {Limousin}, {Cabanac},
  {Marshall}, {Fort}, {Treu}, {P{\'e}llo}, {Jullo}, {Kneib}, \&
  {Sygnet}}]{hong09}
{Tu}, H., {Gavazzi}, R., {Limousin}, M., {et~al.} 2009, \aap, 501, 475

\bibitem[{{van den Bosch} {et~al.}(2008){van den Bosch}, {Pasquali}, {Yang},
  {Mo}, {Weinmann}, {McIntosh}, \& {Aquino}}]{group9}
{van den Bosch}, F.~C., {Pasquali}, A., {Yang}, X., {et~al.} 2008, ArXiv
  e-prints 0805.002

\bibitem[{{Verdugo} {et~al.}(2007){Verdugo}, {de Diego}, \& {Limousin}}]{ver07}
{Verdugo}, T., {de Diego}, J.~A., \& {Limousin}, M. 2007, \apj, 664, 702

\bibitem[{{Willis} {et~al.}(2005){Willis}, {Pacaud}, {Valtchanov}, {Pierre},
  {Ponman}, {Read}, {Andreon}, {Altieri}, {Quintana}, {Dos Santos},
  {Birkinshaw}, {Bremer}, {Duc}, {Galaz}, {Gosset}, {Jones}, \&
  {Surdej}}]{group4}
{Willis}, J.~P., {Pacaud}, F., {Valtchanov}, I., {et~al.} 2005, \mnras, 363,
  675

\bibitem[{{Wilman} {et~al.}(2005){Wilman}, {Balogh}, {Bower}, {Mulchaey},
  {Oemler}, {Carlberg}, {Morris}, \& {Whitaker}}]{Wil05}
{Wilman}, D.~J., {Balogh}, M.~L., {Bower}, R.~G., {et~al.} 2005, \mnras, 358,
  71

\bibitem[{{Yang} {et~al.}(2008){Yang}, {Mo}, \& {van den Bosch}}]{group8}
{Yang}, X., {Mo}, H.~J., \& {van den Bosch}, F.~C. 2008, \apj, 676, 248

\end{thebibliography}

%\newpage

%No por mucho madrugar.
%Al que porfia mata venado
%Para que tanto brinco...

\end{document}